\documentclass[prd,aps,preprintnumbers, nofootinbib,superscriptaddress,notitlepage]{revtex4}
\usepackage{mathrsfs}
\usepackage{amsfonts}
\usepackage{amsmath}
\usepackage{amssymb}
\usepackage{array}
\usepackage{verbatim}
\usepackage{epsfig}
\usepackage{graphicx}
\usepackage{amsbsy}
\usepackage{bm}
\usepackage[dvipsnames]{xcolor}  
\usepackage{graphicx} 
\usepackage[font=small,labelfont=bf]{caption} 
\usepackage[subrefformat=parens]{subcaption}
\usepackage{booktabs}
\usepackage{float}
\usepackage{MnSymbol}

\usepackage{import}
\input{epsf}
\usepackage{psfrag}
\usepackage{xcolor}
\usepackage{slashed}
\usepackage{hyperref}
\usepackage{ulem}
\usepackage{MnSymbol}

\newcommand{\sigmar}{\sigma_r}
\newcommand{\sigmak}{\sigma_k}

\newcommand{\sig}{\sigma}
\newcommand{\hn}{\hat n}
\newcommand{\hr}{\hat r}
\newcommand{\hk}{\hat k}

\newcommand{\rot}{\mathrm{rot}}

\newcommand{\calO}{\mathcal O}

\begin{document}

\begin{flushright}

\end{flushright}

\title{Soft-Radiation-Induced Decoherence of Heavy-Quark Spin Entanglement at the Electron-Ion Collider}


\author{Sanskriti Agrawal}
\affiliation{Department of Physics, Aligarh Muslim University, Aligarh - $202001$, India.}

\author{Muneeb Zahoor}
\affiliation{Department of Physics, Aligarh Muslim University, Aligarh - $202001$, India.}

\author{Raktim Abir}
\affiliation{Department of Physics, Aligarh Muslim University, Aligarh - $202001$, India.}
\email{raktim.ph@amu.ac.in}

\begin{abstract}

Using the soft-gluon theorem, we identify a soft-recoil mechanism by which unresolved gluon radiation induces decoherence in the spin correlations of heavy quark-antiquark pairs produced by longitudinally polarized virtual photons in deep-inelastic scattering. Upon tracing over the unresolved gluon, the reduced spin density matrix is block diagonal in the mirror parity of the 
production plane and is governed by two coefficients: a parity-odd normal-spin flip, which reduces the normal-axis correlation, and a parity-even stochastic rotation of the spin-correlation 
plane about its normal, which suppresses the in-plane spin coherences while leaving the normal-axis correlation unchanged. From the complete tree-level next-to-soft amplitude with the exact 
color-octet structure we derive both coefficients, together with the trace and antisymmetric in-plane terms, in closed form. The parity-odd coefficient is the larger one, by a factor of at 
least $3.5$, and grows as the inverse square of the heavy-quark transverse momentum. We estimate the resulting reduction of concurrence and Bell-CHSH violation, and propose an infrared-safe, 
radiation-binned EIC observable, formed at fixed Born kinematics after longitudinal-transverse separation, which isolates the two channels. We estimate its size, including multiple wide-angle 
emissions in an independent-emission model, the conditions under which detector and kinematic effects cancel, and the statistics required. The latter, however, exceed the yield of exclusive 
heavy-baryon pairs at nominal EIC luminosity by several orders of magnitude, makes the observable elusive for the measurement accessible in that channel. 

\end{abstract}

\maketitle
\section{Introduction}
%
   From Einstein-Podolsky-Rosen's critique questioning the completeness of quantum mechanics \cite{PhysRev.47.777} followed by  Schrödinger's quick rejoinder and identification of entanglement \cite{Schrodinger_1935, Schrodinger_1936} as a defining trait of quantum mechanics in 1935--36, to Bell's/CHSH inequality in the sixties \cite{PhysicsPhysiqueFizika.1.195, PhysRevLett.23.880}, and the first loophole-free Bell tests in 2015 \cite{Hensen:2015ccp, Giustina:2015yza, Stevens:2015awv}, entanglement has provided one of the clearest ways to distinguish genuine quantum correlations from correlations that can be reproduced by classical statistical mixtures in a wide range of physical systems and energy frontiers. 
In recent years, entanglement has begun to play an increasingly important role in high-energy physics as well \cite{Kharzeev:2017qzs}.  A collider event is not merely a dynamical action in phase space contributing to the cross section; it also gives birth to organic quantum correlations between multiple final states \cite{ATLAS:2023fsd}. It can also provide information about the correlations between different degrees of freedom for a single parton \cite{Hatta:2024lbw,Agrawal:2025yoe}.  The spin, momentum, and color correlations of the particles produced in a hard scattering retain detailed information about the underlying quantum nature of the interaction.  When these correlations are reconstructed experimentally, collider processes become laboratories for studying quantum information in relativistic quantum field theory. \\ 

This perspective has gained significant momentum following recent studies of spin correlations in heavy-particle production \cite{Afik:2020onf, Fabbrichesi:2021npl}.  In particular, the observation of spin entanglement in $t{\bar t}$ production at the LHC has demonstrated that genuine quantum correlations can survive in a strongly interacting collider environment and can be extracted from suitable angular observables \cite{ATLAS:2023fsd,CMS:2024pts}.  These developments have opened a new direction at the interface of quantum information theory, particle physics, and quantum chromodynamics.  In this emerging program, QCD radiation is not merely a complication that dilutes a clean partonic signal.  Rather, it is part of the quantum dynamics that determines how entanglement is produced, transported, modified, and, in some cases, degraded. \\ 

Heavy-quark pair production provides a particularly natural setting for this program.  The heavy-quark mass provides a perturbative scale, while the spin correlations of the pair can, in favorable channels, be related to measurable decay or fragmentation products.  At hadron colliders, top quarks are especially clean because they decay before hadronization.  At an Electron-Ion Collider, one can study a complementary class of processes in which a virtual photon probes the gluonic structure of a hadronic target.  Photon-gluon fusion then produces heavy quark-antiquark pairs whose spin density matrix is calculable within perturbative QCD. The EIC, therefore, offers a uniquely clean environment in which one may connect entanglement observables with the dynamics of the quarks and gluons \cite{Qi:2025onf,Fucilla:2025kit,qdb2-k2nh,Fucilla:2026mkg,Liu:2026dzv}.\\

The interface between collider physics and quantum information science provides a new perspective to explore the high energy scattering processes. Significant progress has been made in understanding quantum entanglement in electron-positron collisions \cite{Lin:2025eci,Zhang:2026wvn}, the effects of decoherence in high-energy scattering processes \cite{Aoude:2026eeg,Aoude:2025ovu,Gu:2025RGDecoherence}, and the applications of quantum-information at the Electron-Ion Collider (EIC) \cite{Zhang:2025ean,Robin:2026lqp,Cheng:2025zaw}.\\ 

A significant recent result was obtained by Qi, Guo, and Xiao, who studied spin entanglement and Bell nonlocality in heavy-quark pairs produced through photon-gluon fusion at an EIC \cite{Qi:2025onf}.  They showed that, for a longitudinally polarized virtual photon, the produced quark-antiquark pair is maximally entangled at leading order throughout the kinematic region.  The corresponding spin-correlation matrix describes a pure Bell-type state, and the Bell-CHSH inequality is maximally violated.  For transversely polarized photons, the state is generally mixed, but sizable regions of phase space still exhibit entanglement and Bell nonlocality.  Their analysis identified the longitudinal photon channel as an exceptionally clean leading-order source of maximal spin entanglement in QCD at the EIC. \\ 

A follow-up development was provided by Fucilla and Hatta, in the context of diffractive heavy-quark production, by considering exclusive production through color-singlet Pomeron exchange \cite{Fucilla:2025kit}. It was found that the longitudinal photon again produces a maximally entangled and maximally Bell-violating quark-antiquark pair.  This agreement with the inclusive result is nontrivial, since the Pomeron represents a color-singlet gluonic exchange rather than a single gluon. It suggests that quantum-information observables may provide a new diagnostic of the QCD Pomeron and of gluonic dynamics at high energy.  \\

  This naturally raises a deeper question.  Is this maximal entanglement in the longitudinal channel a robust feature of the underlying QCD dynamics, or is it an idealized leading-order result that naturally degrades once realistic QCD radiations are included? \\ 

This question is the central motivation of the present work.  At leading order, the observed final state consists only of the heavy quark and anti-quark. The spin state of the pair is then pure at the amplitude level, and the longitudinal channel produces maximal entanglement. Beyond leading order, however, real gluon emission contributes. If the emitted gluon is resolved, it becomes part of the measured final state. If it is unresolved, it must be summed over. In the latter case, the experimentally relevant object is then not the pure density matrix of the full quark-antiquark-gluon system, but the reduced spin density matrix of the observed heavy-quark pair, obtained after tracing over the unobserved gluon momentum, polarization, and color.  This tracing operation turns the heavy-quark pair into an open quantum subsystem.   \\

The importance of radiative effects for collider spin entanglement has recently been addressed by Aoude et. al. \cite{Aoude:2025ovu, Aoude:2026eeg}, who formulated unresolved radiation as an open quantum system effect acting on the fermion-pair spin density matrix. In their treatment, leading soft radiation acts trivially on the spin state, while unresolved collinear radiation induces a non-trivial Kraus-map evolution and a calculable reduction of concurrence. This provides an important benchmark for our analysis, which focuses instead on the spin-sensitive subleading soft-gluon recoil mechanism and its role in heavy-quark spin decoherence for the EIC. \\

At this point, the distinction between ordinary radiative corrections and genuine spin decoherence is essential. In the case of soft radiation, the leading soft-gluon factor is eikonal and spin independent.  It changes the inclusive rate and contributes to the familiar infrared structure of QCD but, after normalization of the spin-density matrix, does not rotate or distort the leading-order spin-correlation pattern. Decoherence requires a more subtle mechanism: unresolved radiation must carry information about the spin geometry of the hard scattering.  Such information first appears through subleading soft effects, where recoil and angular momentum enter the amplitude.  These terms are less singular than the leading eikonal factor, but they are precisely the terms capable of changing the normalized spin density matrix.\\ 

In this paper, we show that, for the longitudinal photon, unresolved soft gluon radiation decoheres the pair through two distinct channels.  At the leading order, the spin operator of the heavy-quark pair lies in the production plane, and the correlation along the normal to this plane is maximal.  At next to leading order, the spin-dependent soft recoil has a component odd under reflection through the production plane, which flips the spin normal to the plane and reduces the normal-axis correlation.  It also has a component even under this reflection, which produces a small, event-by-event rotation of the spin-correlation plane about its normal.  For a gluon whose momentum, polarization and color were all measured, both displacements would be coherent and unitary.  When the gluon is unresolved, they are averaged over the soft phase space. The mean rotation acts coherently and does not decohere the state, but the spread of the rotation angle survives the average and opens a dephasing channel that suppresses the in-plane spin correlations. The normal-spin flip is the larger of the two effects throughout the kinematic range, so that the normal-axis correlation is degraded more strongly than the in-plane coherence. Consequently, the initially maximally entangled state slips away from maximal entanglement, with a characteristic anisotropy.
In this framework, the unresolved gluon acts as an environment. It encodes information about recoil and about the orientation of the heavy-quark spin-correlation plane. Since that information is not measured, it is inaccessible to the observed subsystem. The full final state remains quantum mechanical and evolves unitarily, but the reduced state of the heavy-quark pair becomes mixed.  \\

The paper is organized as follows. We first review the leading-order spin structure of the longitudinal photon channel and the origin of maximal heavy-quark spin entanglement \cite{Qi:2025onf, Fucilla:2025kit}. We then use the subleading soft-gluon theorem to identify how unresolved radiation induces recoil-driven normal-spin flips and rotations of the spin-correlation plane. We obtain the two decoherence rates in closed form and discuss, within an independent-emission model, how multiple unresolved emissions lead to an exponential suppression of the normal-axis correlation and of the in-plane spin coherences. Finally, we estimate the resulting reduction of entanglement and Bell violation and propose a radiation-binned Electron-Ion Collider observable that isolates this anisotropic pattern of soft-radiation-induced decoherence.

\vspace{0.25cm} 

\section{Leading Order spin structure}

    In this study, we consider the process $\gamma^*(q)+g(p)\rightarrow q(p_1)+{\bar{q}}(p_2)+g(k)$ where the final state soft gluon is unresolved. We work in the center-of-mass frame of the observed heavy quark-antiquark pair in the helicity basis $\{\hat{n},\hat{r}, \hat{k}\}$, where $\hat{k}$ is the direction of the outgoing quark, $\hat{n}$ is orthogonal to the production plane spanned by the incoming virtual photon direction and the outgoing quark momentum, $\hat{r}$ is orthogonal to $\hat{k}$ and $\hat{n}$. The ordered set $\{\hat{n},\hat{r}, \hat{k}\}$ is an orthonormal basis where $\hat{k}=\left(\sin\theta\cos\varphi, \sin\theta\sin\varphi, \cos\theta\right)$, with $\varphi$ the quark azimuth about the photon axis. We define the invariant mass of $q{\bar q}$ pair by $M^2=\left(p_1+p_2\right)^2$, and also define the heavy-quark velocity, $  \beta=\sqrt{1-{4m^2}/{M^2}}$ where $m$ is the mass of the heavy quarks. The angle $\theta$ represents the angle between the virtual-photon direction and the quark momentum. As we work in the heavy-quark-pair center-of-mass frame, the observed heavy quark and antiquark momenta are
\begin{equation}
  p_1^\mu=E(1,\beta\hk),\qquad
  p_2^\mu=E(1,-\beta\hk),
  \qquad E=\frac{M}{2},\qquad
  \beta=\sqrt{1-\frac{4m^2}{M^2}} .
\end{equation}
Let the virtual photon define the longitudinal direction. The orthonormal triad is
\begin{equation}
  \hk=(\sin\theta,0,\cos\theta),\qquad
  \hr=(\cos\theta,0,-\sin\theta),\qquad
  \hn=(0,1,0),
\end{equation}
where \(\hr\) lies in the production plane and \(\hn\) is normal to it. Equivalently, $\hn=\hat z\times\hk/\sin\theta$ and $\hr=\hn\times\hk$, so that $\{\hk,\hr,\hn\}$ is a right-handed triad, while the ordering $\{\hn,\hr,\hk\}$ used for matrices below is left-handed. For the Pauli matrices $\sigma_a\equiv\hat a\cdot\boldsymbol\sigma$ this gives
\begin{equation}
  [\sigma_n,\sigma_r]=-2i\sigma_k,\qquad
  [\sigma_n,\sigma_k]=2i\sigma_r,\qquad
  \sigma_r\sigma_k=-i\sigma_n ,
  \label{eq:pauli-triad}
\end{equation}
which we use throughout. The quark spin is quantized along $\{\hn,\hr,\hk\}$ and the antiquark spin along $\{\hn,-\hr,-\hk\}$; the spin-correlation matrix $C_{ij}$ below refers to these axes. A generic soft-gluon direction relative to the outgoing heavy quark is parametrized by a polar angle $\chi$ and an azimuth $\phi_g$ about $\hk$,
\begin{equation}
  \hat \ell
  =
  \cos\chi\,\hk
  +\sin\chi\cos\phi_g\,\hr
  +\sin\chi\sin\phi_g\,\hn,
  \qquad k^\mu=\omega(1,\hat \ell).
\end{equation}
Thus,
\begin{equation}
  p_1\cdot k=E\omega(1-\beta\cos\chi),
  \label{eq:p1dotk}
\end{equation}
which will eventually be the universal soft-recoil denominator.
\\
    
    \noindent
In the leading order there is no final state gluon. It is therefore $2\rightarrow2$ process as $\gamma^*g\rightarrow q{\bar q}$. The Dirac spinors in the Pauli-spinor decomposition for the $q{\bar q}$ pair produced in the CM frame are defined as,
\begin{eqnarray}
     u(p_1,\zeta)=
     \begin{pmatrix}
     a\zeta\\
     b\sigma_k \zeta
     \end{pmatrix},
     ~~~~~~~~~~~ v(p_2,\eta)=\begin{pmatrix}
         -b\sigma_k \eta\\
         a\eta
     \end{pmatrix},
\end{eqnarray}
where, $a,b=\sqrt{E\pm m}$, $E=M/2$ and, $\zeta$ (and $\eta$) are the Pauli quark (anti-quark) spinors. The spin amplitude can be written as,
\begin{eqnarray} 
    {\cal M}_{\lambda_q \lambda_{\bar q}} \sim{\bar v}(p_2)\gamma^+u(p_1)
    =\eta^{\dagger}_{\lambda_{\bar q}} X_0 ~\zeta_{\lambda_q},
\end{eqnarray}
    where $\gamma^+$ stands schematically for the Dirac structure of the longitudinal-photon Born amplitude, and $X_0$ is the spin-operator, acting on the Pauli-spin space, encodes the spin structure of the interaction process. The relation between $X_0$ and the two-spin wave function is given in Appendix~A. For the longitudinal photon channel, in the helicity basis the leading order spin-amplitude can be represented by the spin operator,
\begin{equation}
  X_{0}={ A}(\beta,\theta)\,\sigmar+{ B}(\beta,\theta)\,\sigmak.
  \label{eq:lo_operator}
\end{equation}
Here $\sigma_r,\sigma_k$ are the Pauli matrix projected along the basis directions $\{\hat{r},\hat{k}\}$ and the functions ${ A}$ and ${ B}$ are real kinematic coefficients depending upon the quark velocity and the scattering angle $\theta$.  Although their overall normalization is irrelevant for the normalized spin-correlation matrix, their relative size determines the orientation of the spin operator $X_0$ in the $\{\hat{r},\hat{k}\}$ plane.
One may notice the absence of the identity $I$ component as well as the $\sigma_n$ component in Eq.\eqref{eq:lo_operator}. Therefore, the longitudinally polarized virtual photon produces a spin operator which lies entirely in the $\{\hat{r},\hat{k}\}$ plane or the production plane. This operator constrains the spin state of the $q{\bar q}$ pair produced in the interaction.\\

It is convenient to describe the two-particle spin state in the density matrix formalism to explicitly show the spin correlation.  To obtain the density matrix for a process, we take the square of the scattering amplitude, sum over the spin of the initial states and keep the spinor indices open for the final state.  For the longitudinal case, the amplitude squared is proportional to
\begin{eqnarray}
    {\bar u}_{\beta'}(p_1)\gamma^+v_{\beta}(p_2){\bar v_{\alpha}(p_2)}\gamma^+u_{\alpha'}(p_1).
\end{eqnarray}
The spin density matrix can further be decomposed into $4 \times 4$ Hermitian matrix as,
\begin{equation}
  \rho=\frac{1}{4}\left[I\otimes I+ P_i \sigma_i\otimes I+\bar{ P}_jI\otimes\sigma_j+
  \sum_{i,j}C_{ij}\,\sigma_i\otimes\sigma_j\right],
  \label{eq:density_general}
\end{equation}
where $ P$ and $\bar{ P}$ are the particle polarization vectors of the quark and anti-quark, and $C_{ij}$ is the spin-correlation matrix. The normalized longitudinal spin-correlation matrix at leading order written in the basis $\{\hat{n},\hat{r},\hat{k}\}$ is found to be the form \cite{Qi:2025onf},
\begin{equation}
C^L_{\rm LO}= 
\begin{pmatrix}
1&0&0\\
0&C_{rr}&C_{rk}\\
0&C_{kr}&C_{kk}
\end{pmatrix},
\label{eq:clo}
\end{equation}
where, the non-zero components are given by,
\begin{align}
 C_{rr}=-C_{kk}
 &=\frac{1-(2-\beta^2)\cos^2\theta}{1-\beta^2\cos^2\theta},
 \label{eq:crr}\\
 C_{rk}=C_{kr}
 &=-\frac{\sqrt{1-\beta^2}\,\sin 2\theta}{1-\beta^2\cos^2\theta}.
 \label{eq:crk}
\end{align}
The structure of the $C$-matrix reflects the planar nature of the longitudinally polarized photon interaction with the gluon.
While the component $C_{nn}=1$ indicates perfect correlation along the normal direction, the remaining $2\times2$ block shows the nontrivial correlation in the production plane. The denominator $1-\beta^2\cos^2\theta$
coming from the normalized squared amplitude, governs the angular and velocity dependence of the recoil. With the map of Appendix~A, Eqs.~\eqref{eq:crr}--\eqref{eq:crk} correspond to
\begin{equation}
A=\sin\theta,\qquad B=-\sqrt{1-\beta^2}\,\cos\theta ,
\label{eq:AB}
\end{equation}
up to a common normalization, for which the in-plane block is $C_\parallel=2\hat{\mathbf x}\hat{\mathbf x}^T-I$, with $\hat{\mathbf x}=(A,B)/\sqrt{A^2+B^2}$ in the $\{\hr,\hk\}$ components and $A^2+B^2=1-\beta^2\cos^2\theta$. \\

\section{Soft Emission and Universal Soft Theorem}
\label{sec:nlo-real-soft}

The $C$-matrix in Eq.~\eqref{eq:clo} represents a pure maximally entangled spin state. At the leading order, the spin directions of the produced $q{\bar q}$ pair are correlated in a way that cannot be reproduced by a classical mixture of independently polarized quarks. At this order there is no final-state radiation, so no information is lost from the $q{\bar q}$ system and the density matrix is pure. When a real gluon is emitted, the final state lives in the larger Hilbert space ${\cal H}_{q}\otimes {\cal H}_{\bar q}\otimes {\cal H}_{g}$. If the gluon is not observed, the relevant object is the reduced density matrix $\rho_{q\bar q}^{\rm red}=\operatorname{Tr}_g\,\rho_{q\bar q g}$, where $\operatorname{Tr}_g$ denotes a sum over the gluon polarization and color, together with an integral over the unresolved momentum region. This partial trace is the origin of decoherence whenever the spin state of the pair depends on the state of the unobserved gluon. \\

\subsection{Soft expansion of the real-emission amplitude}

Throughout we work at a fixed, resolved Born phase-space point $\Phi_B$, specified by the observed pair variables $(M,\beta,\theta)$, by $Q^2$, and by the photon direction in the pair rest frame, which together fix the Born spin operator $X_0$ of Eq.~\eqref{eq:lo_operator}. A real gluon is labelled by $\xi=(k,\lambda,a)$, with $k^\mu=\omega(1,\hat\ell)$, physical polarization $\lambda$, and adjoint color $a$. Keeping the heavy-quark spin indices open, the real-emission amplitude is a $2\times2$ Pauli operator which, for every $\xi$, we decompose as
\begin{equation}
X_g^{\lambda a}(k)=\delta c_0\, I+\delta c_n\,\sigma_n+\delta c_r\,\sigma_r+\delta c_k\,\sigma_k .
\label{eq:Xg-decomp}
\end{equation}
For emission of a soft gluon the soft theorem gives
\begin{equation}
  \mathcal M_{q{\bar q} g}^a(k) = g_s\sum_i \mathbf T_i^a
  \left[ \frac{p_i\cdot\epsilon^*}{p_i\cdot k} + \frac{\epsilon^*_\mu k_\nu J_i^{\mu\nu}}{p_i\cdot k} \right] \mathcal M_{q{\bar q}}+O(\omega) ,
  \label{eq:soft-theorem}
\end{equation}
where the sum runs over the quark, the antiquark and the incoming gluon, with the color charges $\mathbf T_i$ written in the all-outgoing convention, $\mathbf T_q+\mathbf T_{\bar q}+\mathbf T_g=0$. The factorization of amplitudes in the soft gluon limit goes back to the classic soft-photon theorems of Low and Burnett-Kroll, and to Weinberg's formulation of universal infrared factorization. In QCD, the leading soft-gluon factor is the non-Abelian eikonal current, while the first subleading, or next-to-soft, term is the gauge-theory analogue of the Low-Burnett-Kroll theorem and involves the angular-momentum operator acting on the hard amplitude \cite{Low:1958sn, Burnett:1967km,Weinberg:1965nx, Catani:2000pi, DelDuca:1990gz, Casali:2014xpa, Bern:2014vva}. The angular-momentum generator decomposes into orbital and intrinsic-spin parts,
\begin{equation}
  J_i^{\mu\nu}=L_i^{\mu\nu}+S_i^{\mu\nu}, \qquad
  L_i^{\mu\nu} = p_i^\mu\frac{\partial}{\partial p_{i\nu}} -p_i^\nu\frac{\partial}{\partial p_{i\mu}} .
  \label{eq:Lmunu}
\end{equation}
It is convenient to make the power counting explicit by writing Eq.~\eqref{eq:soft-theorem} as a Laurent expansion in the gluon energy,
\begin{equation}
X_g^{\lambda a}(k)=g_s\left[\frac{1}{\omega}\,m_{-1}^{\lambda a}(\hat\ell)+m_{0}^{\lambda a}(\hat\ell)+O(\omega)\right],
\label{eq:laurent}
\end{equation}
where $m_{-1}/\omega$ is the eikonal term and $m_0$ the next-to-soft term. Since $p_i\cdot\epsilon^*/p_i\cdot k\sim 1/\omega$ while $\epsilon^*_\mu k_\nu J_i^{\mu\nu}/p_i\cdot k\sim\omega^0$ and the derivatives in $L_i$ act on a hard amplitude whose only scale is $E=M/2$, one has $m_0/m_{-1}=O(1/E)$.

\paragraph*{Eikonal term.} The eikonal term is a scalar in the final-state spin space: it multiplies the leading-order Pauli operator $X_0$ by a color- and polarization-dependent number, $m_{-1}^{\lambda a}=c_{-1}^{\lambda a}(\hat\ell)\,X_0$. Consequently it changes the real-emission rate but not the normalized spin state, and it is not the origin of decoherence. The Born pair in $\gamma^*g\to q\bar q$ is a color octet, so that $\mathbf T_q\cdot\mathbf T_{\bar q}=(C_A-2C_F)/2=1/(2N_c)$ and $\mathbf T_q\cdot\mathbf T_g=\mathbf T_{\bar q}\cdot\mathbf T_g=-C_A/2$. After the physical polarization sum, the color-weighted eikonal kernel is therefore
\begin{equation}
W_0(k)\equiv-\sum_{i,j}\mathbf T_i\cdot\mathbf T_j\,\frac{p_i\cdot p_j}{(p_i\cdot k)(p_j\cdot k)}
=C_A\left(w_{qg}+w_{\bar q g}\right)-\frac{1}{N_c}\,w_{q\bar q}-C_F\left(w_{qq}+w_{\bar q\bar q}\right),
\qquad w_{ij}=\frac{p_i\cdot p_j}{(p_i\cdot k)(p_j\cdot k)} ,
\label{eq:W0}
\end{equation}
whose closed form in the pair rest frame is given in Appendix~A. The initial-final dipoles carry the largest color coefficients, so no single effective color factor describes the emission pattern. Equation~\eqref{eq:W0} is reproduced, direction by direction, by the complete tree-level amplitude of Appendix~B.

\paragraph*{Next-to-soft term.} The orbital part of the next-to-soft operator acts as a differential displacement of the hard momenta,
\begin{equation}
\frac{\epsilon_\mu^\ast k_\nu L_i^{\mu\nu}}{p_i\cdot k} = \delta p_{i,\alpha}^{(\epsilon,k)}\,\frac{\partial}{\partial p_{i\alpha}},
\qquad
\delta p_{i,\alpha}^{(\epsilon,k)} \equiv \frac{p_i\cdot \epsilon^\ast}{p_i\cdot k}\, k_\alpha - \epsilon_\alpha^\ast ,
\qquad p_i^\alpha\,\delta p_{i,\alpha}^{(\epsilon,k)}=0 ,
\label{eq:orbital_displacement_operator}
\end{equation}
so that it changes the direction of the hard momentum without changing its mass. Since $X_0$ depends on the hard momenta through $\beta$, $\theta$ and the basis matrices $\sigma_r,\sigma_k$, the chain rule gives
\begin{equation}
\delta X_{{\rm orb},i}
=
g_s\mathbf T_i^a \Big[ \delta\beta_i\,\partial_\beta X_0 + \delta\theta_i\,\partial_\theta X_0 +
A\,\delta\sigma_r + B\,\delta\sigma_k \Big],
\label{eq:delta_X_orb_chain_rule}
\end{equation}
with $\delta\beta_i=\delta p_{i,\alpha}\partial\beta/\partial p_{i\alpha}$ and similarly for $\theta$ and for the basis matrices. The intrinsic-spin part $S_i^{\mu\nu}$ acts on the spinor indices. Both parts are of the same order in $\omega/E$.

\subsection{Next-to-soft spin displacement and power counting}
\label{sec:powercount}

Because the eikonal amplitude is proportional to $X_0$, the normalized spin operator of an emitted state with fixed $\xi$ is
\begin{equation}
\widehat X_\xi=X_0+\frac{\omega}{c_{-1}^{\lambda a}(\hat\ell)}\,m_0^{\lambda a}(\hat\ell)+O\!\left(\frac{\omega^2}{E^2}\right),
\label{eq:Xhat}
\end{equation}
up to normalization. The displacement of the emitted state away from the Born state is the component of $\omega\,m_0/c_{-1}$ orthogonal to $X_0$ in the Hilbert-Schmidt inner product. For $X_0=A\sigma_r+B\sigma_k$ the orthogonal complement is spanned by
\begin{equation}
T_n=A\sigma_k-B\sigma_r,\qquad I,\qquad \sigma_n ,
\label{eq:complement}
\end{equation}
with ${\rm Tr}(T_n^\dagger X_0)=0$ and ${\rm Tr}(T_n^\dagger T_n)=2(A^2+B^2)$. The rotation angle of the spin-correlation plane is defined as a ratio of amplitudes,
\begin{equation}
\delta\phi_n(\xi)\equiv\frac{\omega}{c_{-1}^{\lambda a}(\hat\ell)}\,
\frac{{\rm Tr}\left[T_n^\dagger\,m_0^{\lambda a}(\hat\ell)\right]}{{\rm Tr}\left[T_n^\dagger T_n\right]}
=\frac{\omega}{E}\,f_n(\hat\ell;\lambda,a;\beta,\theta)+O\!\left(\frac{\omega^2}{E^2}\right),
\label{eq:dphi-ratio}
\end{equation}
where $f_n$ is a dimensionless function of the gluon direction, of its color and polarization labels, and of the Born kinematics. The components along $I$ and $\sigma_n$ are defined in the same way and are of the same order. 

The leading emission probability per unit energy is $\propto d\omega/\omega$, and the squared displacement relative to the emitted state is $\propto(\omega/E)^2$, so that
\begin{equation}
\frac{d\omega}{\omega}\left(\frac{\omega}{E}\right)^2=\frac{\omega\,d\omega}{E^2}.
\label{eq:measure}
\end{equation}
Equivalently, the decohering contribution is $|m_0|^2$ integrated with the flat measure $\omega\,d\omega$; the eikonal factor $1/\omega^2$ appears only in the normalization of the emitted state. 

\subsection{Normalized density-matrix map}
\label{sec:map}

We first collect the assumptions on which the map rests. (i) The Born kinematics $\Phi_B$ is held fixed and the photon is longitudinal; in an experiment this requires the longitudinal-transverse separation and fixed-Born construction of Sec.~\ref{sec:eic_observable}. (ii) At most one real gluon is emitted, and the angular region in which it is unresolved is symmetric under the reflection $\mathcal R_n$ through the production plane (Sec.~\ref{sec:parity}); this is the ${\cal O}(\alpha_s)$ fixed-order content. Multiple emissions are treated separately, in an independent-emission model, in Secs.~\ref{sec:multiple} and \ref{sec:multi-bins}. (iii) The emission is soft, and only the leading power in $\tau_c=E_{\rm cut}/E$ and in $E_{\rm cut}/p_T$ is kept, Eq.~\eqref{eq:window} below. (iv) The amplitudes are tree level, so there are no absorptive phases. (v) Hard virtual corrections and radiation collinear to the beam or to the heavy quarks do not depend on the wide-angle resolution at leading power, and they cancel in the radiation-binned ratios of Sec.~\ref{sec:eic_observable} up to the residual terms listed in Sec.~\ref{sec:collinear}. The fixed-order map below, and its consequences in Secs.~\ref{sec:dephasing} and \ref{sec:measures}, follow from (i)--(v) alone. The exponentiated form of Sec.~\ref{sec:multiple} requires in addition that successive emissions are statistically independent.

At ${\cal O}(\alpha_s)$ the measured, unnormalized spin matrix at fixed $\Phi_B$ is the sum of the Born term $B_0=\sigma_0\rho^{(0)}$ with ${\rm Tr}\rho^{(0)}=1$, the virtual correction $V$ (including the initial-state collinear counterterm), and the real emission $R_{\cal U}$ integrated over the unresolved region ${\cal U}$ defined by the measurement,
\begin{equation}
\rho_{\cal U}=\frac{B_0+V+R_{\cal U}}{{\rm Tr}\left(B_0+V+R_{\cal U}\right)}
=\rho^{(0)}+\frac{1}{\sigma_0}\Big[\Delta B-\rho^{(0)}\,{\rm Tr}\,\Delta B\Big]+O(\alpha_s^2),
\qquad \Delta B=V+R_{\cal U}.
\label{eq:map}
\end{equation}
The subtraction is required by normalization and makes the map trace preserving order by order. Any contribution proportional to the Born matrix, $\Delta B=c\,\sigma_0\rho^{(0)}$, drops out of Eq.~\eqref{eq:map}. This removes the eikonal real emission and the spin-independent infrared part of the virtual correction, whose singularities cancel against each other in the rate as usual. The finite spin-dependent part of the hard virtual correction does not depend on the radiation resolution and is common to all radiation bins considered in Sec.~\ref{sec:eic_observable}.

Using Eq.~\eqref{eq:laurent}, the real-emission matrix contains three terms. The eikonal term is proportional to $\rho^{(0)}$ and cancels. The interference between $m_{-1}$ and $m_0$ is linear in the displacement and integrates with the measure $d\omega$; it produces a coherent, ${\cal O}(\alpha_s E_{\rm cut}/E)$ rotation of the spin state, which changes $C_{rr}$ and $C_{rk}$ individually but, being a first-order deformation of a pure state, changes neither the purity nor the entanglement of the maximally entangled Born state, nor any of the rotation-invariant correlations used below. The first decohering contribution is the quadratic term. Defining the unit directions $\widehat T_A\in\{T_n/\sqrt{A^2+B^2},\,I,\,\sigma_n\}$ in the complement \eqref{eq:complement}, it is the positive (Gram) matrix
\begin{equation}
\Sigma_{AB}=\frac{1}{{\cal N}_B}\int_{\cal U}d\Phi_g\,g_s^2\sum_{\lambda,a}
\frac{{\rm Tr}\big[\widehat T_A^\dagger m_0^{\lambda a}\big]\,{\rm Tr}\big[m_0^{\lambda a\,\dagger}\widehat T_B\big]}{4},
\qquad
z_A^*\Sigma_{AB}z_B\ge0 ,
\label{eq:Sigma}
\end{equation}
where ${\cal N}_B$ is the color- and spin-summed Born norm with the same initial-state averages. Here and below we use the inner product $\langle X|Y\rangle\equiv\frac12{\rm Tr}(X^\dagger Y)$ on Pauli operators, which equals the overlap of the corresponding two-spin states (Appendix~A). The directions $\widehat T_A$ are orthonormal in it, the factor $1/4$ in Eq.~\eqref{eq:Sigma} is the product of the two factors $\frac12$, and ${\cal N}_B=\sum\langle X_0|X_0\rangle$ with the Born color factor. The explicit chain from the amplitude to $\Sigma_{AB}$ is written out in Eq.~\eqref{eq:chain} of Appendix~A. With $g_s^2\,d^3k/[(2\pi)^32\omega]=(\alpha_s/\pi)\,\omega\,d\omega\,d\Omega/(4\pi)$ and a cumulative energy resolution $\omega<E_{\rm cut}$,
\begin{equation}
\Sigma_{AB}(\tau_c)=\frac{\alpha_s}{2\pi}\,\tau_c^2\,K_{AB}(\beta,\theta),
\qquad \tau_c\equiv\frac{E_{\rm cut}}{E},
\qquad
K_{AB}=E^2\int_{\Omega}\frac{d\Omega}{4\pi}\,\frac{1}{{\cal N}_B}\sum_{\lambda,a}\frac{{\rm Tr}\big[\widehat T_A^\dagger m_0\big]{\rm Tr}\big[m_0^{\dagger}\widehat T_B\big]}{4},
\label{eq:Kdef}
\end{equation}
where $\Omega$ is the angular region in which the radiation is unresolved. We write $K_{AB}$ for the full solid angle and $K^{\rm wa}_{AB}$ for the wide-angle region $\Omega_{\rm wa}$ defined by the measurement in Sec.~\ref{sec:eic_observable}. The decohering part of the inclusive map is then
\begin{equation}
\delta\rho=\sum_{A,B}\Sigma_{AB}\Big(|\widehat T_A\rangle\langle \widehat T_B|-\delta_{AB}\,\rho^{(0)}\Big),
\label{eq:drho}
\end{equation}
where $|\widehat T_A\rangle$ is the two-spin state associated with the Pauli operator $\widehat T_A$. Equations~\eqref{eq:map}--\eqref{eq:drho} include the no-emission sector and the virtual correction through the normalization, and therefore define a trace-preserving, consistently normalized density-matrix map at this order.

\paragraph*{Conditional versus inclusive averages.} Equation~\eqref{eq:drho} is the contribution of the unresolved region to the inclusive, unconditioned state; it carries an overall factor $\alpha_s$. A different object is the state conditioned on the presence of one real emission inside a shell $\omega_a<\omega<\omega_b$. There the overall factor $g_s^2$ appears in both the numerator and the denominator and cancels, and the normalization is the eikonal emission probability. This is the state of the higher-radiation bin of Sec.~\ref{sec:resolution}. To make the structure explicit, write the normalized emitted state for fixed $\xi$ as $|\psi_\xi\rangle\propto|B\rangle+\sum_A d_A(\xi)\,|\widehat T_A\rangle$, with $|B\rangle$ the normalized Born state and $d_A=\langle\widehat T_A|\,\omega\,m_0/c_{-1}\rangle/\langle X_0|X_0\rangle^{1/2}={\cal O}(\omega/E)$. Averaging $|\psi_\xi\rangle\langle\psi_\xi|$ with the normalized eikonal weight of the shell and expanding to second order in $d$,
\begin{equation}
\rho^{\rm cond}=|B'\rangle\langle B'|
+\sum_{A,B}\Sigma^{\rm cond}_{AB}\Big(|\widehat T_A\rangle\langle\widehat T_B|-\delta_{AB}\,\rho^{(0)}\Big)+{\cal O}(d^3),
\qquad
\Sigma^{\rm cond}_{AB}=\langle d_Ad_B^*\rangle-\langle d_A\rangle\langle d_B\rangle^* ,
\label{eq:cond-cov}
\end{equation}
where $|B'\rangle\propto|B\rangle+\sum_A\langle d_A\rangle|\widehat T_A\rangle$ is the coherently displaced pure state. The decohering part is therefore the covariance of the displacement, not its second moment. By the reflection symmetry of Sec.~\ref{sec:parity}, $\langle d_A\rangle=0$ in the mirror-odd sector, so the mean enters only the rotation channel. The second moment follows from the energy integrals of Appendix~A,
\begin{equation}
\langle d_Ad_B^*\rangle[\omega_a,\omega_b]=\frac{\omega_b^2-\omega_a^2}{2E^2\,\ln(\omega_b/\omega_a)}\,\frac{K_{AB}}{\overline W_0}
+O\!\left(\frac{\omega_b^3}{E^3}\right),
\qquad
\overline W_0=\int_{\Omega}\frac{d\Omega}{4\pi}\,\omega^2 W_0 .
\label{eq:Sigma-cond}
\end{equation}
For the inclusive state the same decomposition holds, but the mean displacement is ${\cal O}(\alpha_s\tau_c)$ and its square is ${\cal O}(\alpha_s^2)$, so that Eq.~\eqref{eq:drho} is the covariance to the order considered. The two averages are physically distinct, and we keep them separate below.

\section{Infinitesimal rotation of the leading spin operator}
\label{sec:orbital-only-clean}

The leading operator $X_0=A\sig_r+B\sig_k$ may be viewed as a vector in the two-dimensional plane spanned by $\sig_r$ and $\sig_k$. A change of the observed axes $\{\hat r,\hat k\}$, at fixed physical state, changes the matrices $\sigma_r,\sigma_k$ but not the state; it is a passive relabeling and carries no physical information. A physical (active) rotation of the spin state by an angle $\delta\phi_n$ about $\hn$, at fixed axes, is
\begin{equation}
X_0 \to e^{-i\delta\phi_n\sig_n/2}~X_0~e^{+i\delta\phi_n\sig_n/2},
\qquad
\delta X_{\rot}=\frac{i}{2}\delta\phi_n[\sig_n,X_0]=\delta\phi_n\,(A\sigmak-B\sigmar)\equiv\delta\phi_n\,T_n ,
\label{eq:active}
\end{equation}
where we used the commutators \eqref{eq:pauli-triad}. Only the difference between the state produced in the radiative event and the Born state at the same observed kinematics, both expressed in the same observed axes, is physical. This difference is what Eqs.~\eqref{eq:Xhat} and \eqref{eq:dphi-ratio} measure. The overall sign of $\delta\phi_n$ depends on the orientation convention; the dephasing depends only on its second moments.

\subsection{Mirror parity of the production plane and the two channels}
\label{sec:parity}

The reflection $\mathcal R_n$ through the production plane is a symmetry of the Born configuration: the photon, the incoming gluon and the heavy-quark momenta lie in the plane. On the two spins, $\mathcal R_n$ (parity followed by a rotation by $\pi$ about $\hat n$) acts as $(-i\sigma_n)\otimes(-i\sigma_n)=-\Pi_n$, with
\begin{equation}
\Pi_n\equiv\sigma_n\otimes\sigma_n ,\qquad \Pi_n^2=1 .
\label{eq:Pi}
\end{equation}
The longitudinal Born state is an eigenstate of $\Pi_n$ with eigenvalue $+1$: only the in-plane gluon polarization contributes at Born level, and $X_0$ lies in the $(\sigma_r,\sigma_k)$ plane. This is the symmetry reason for $C_{nn}={\rm Tr}[\rho\,\Pi_n]=1$ in Eq.~\eqref{eq:clo}. In the Pauli language, the $\Pi_n=+1$ sector is spanned by $\{X_0,T_n\}$ and the $\Pi_n=-1$ sector by $\{I,\sigma_n\}$.

A soft gluon emitted at a generic angle, or with a polarization normal to the plane, breaks the reflection symmetry event by event. The real-emission amplitude then has components in both sectors. Since the amplitude is covariant under $\mathcal R_n$, for a mirror-even initial state the pair is in the $\Pi_n=+1$ ($-1$) sector when the emitted gluon is in a mirror-even (mirror-odd) state; the normal polarization of the incoming gluon, which is forbidden at Born level, reverses this correspondence. A $\Pi_n=+1$ displacement is necessarily along $T_n$: it is a rotation of the spin-correlation plane about $\hat n$. A $\Pi_n=-1$ admixture lies along $I$ and $\sigma_n$: it flips the normal-spin parity of the pair. The initial-state polarizations are summed incoherently, and for a reflection-symmetric measurement region the overlap of mirror-even and mirror-odd gluon states vanishes after integration. The trace over the gluon therefore leaves no interference between the sectors, so that
\begin{equation}
\rho^{\rm red}=\rho_{+}\oplus\rho_{-},\qquad
C_{nn}={\rm Tr}\big[\rho^{\rm red}\,\Pi_n\big]=1-2P_{-},\qquad
C_{nr}=C_{nk}=C_{rn}=C_{kn}=0,
\label{eq:blockdiag}
\end{equation}
where $P_-$ is the total probability of the $\Pi_n=-1$ sector. Equation~\eqref{eq:blockdiag} is exact at the order considered and shows that the normal correlation is protected only within the mirror-even (rotation) sector. The two coefficients that govern the map are
\begin{equation}
\Gamma_n\equiv2\,\Sigma_{TT}=\frac{\alpha_s}{\pi}\,\tau_c^2\,K_n(\beta,\theta),
\qquad
\Gamma_\perp\equiv2P_-=2\left(\Sigma_{II}+\Sigma_{nn}\right)=\frac{\alpha_s}{\pi}\,\tau_c^2\,K_\perp(\beta,\theta),
\label{eq:Gammas}
\end{equation}
with $K_n\equiv K_{TT}$ and $K_\perp\equiv K_{II}+K_{nn}$ from Eq.~\eqref{eq:Kdef}. Since the displacement along $\widehat T_n$ is $\delta\phi_n$ of Eq.~\eqref{eq:dphi-ratio}, one has $\Gamma_n=2\langle\delta\phi_n^2\rangle$, where the average is the second moment with the inclusive, emission-weighted measure of Eq.~\eqref{eq:Sigma}. For the inclusive state this coincides with twice the variance to the order considered, since $\langle\delta\phi_n\rangle^2={\cal O}(\alpha_s^2)$. For the conditional state of Eq.~\eqref{eq:cond-cov} the variance, $\langle\delta\phi_n^2\rangle-\langle\delta\phi_n\rangle^2$, is the relevant quantity.

\subsection{In-plane recoil from the angular shift}

This and the following two subsections give a heuristic decomposition of the next-to-soft displacement into recoil and intrinsic-spin pieces. It explains the parametric behavior of the kernels, but the kernels themselves, Sec.~\ref{sec:kernels}, are computed from the complete amplitude and do not rely on it. The leading spin operator is not only a vector in the $(\sigma_r,\sigma_k)$ operator plane; it is also a function of the Born scattering angle, $X_0(\beta,\theta)=A(\beta,\theta)\sigma_r+B(\beta,\theta)\sigma_k$. A soft recoil that shifts the hard angle, $\theta\to\theta+\delta\theta$, induces the variation
\begin{equation}
\delta X_0^{(\theta)} = \delta\theta \left[ (A_\theta+B)\sigma_r + (B_\theta-A)\sigma_k \right] + \delta\varphi
\left[ A\cos\theta+B\sin\theta \right]\sigma_n ,
\label{eq:delta_X0_theta_collect}
\end{equation}
where $A_\theta\equiv\partial_\theta A$, $B_\theta\equiv\partial_\theta B$, and $\delta\varphi$ is the azimuthal recoil of the quark direction about the photon axis (the intermediate Pauli algebra is given in Appendix~A). The projection onto $T_n$ contains a term $-\delta\theta$ from $A\,\delta\sigma_r+B\,\delta\sigma_k$, i.e.\ from the transport of the spin axes with the quark direction. Whether this term is attributed to the orbital operator or to the intrinsic-spin operator depends on how the external spinors are transported, Sec.~\ref{sec:intrinsic}, and it is not separately meaningful. The part of the in-plane motion that is independent of this choice is the connection term
\begin{equation}
\delta\phi_n^{(\theta)}
=
\frac{AB_\theta-BA_\theta}{A^2+B^2}\,\delta\theta
=\frac{\sqrt{1-\beta^2}} {1-\beta^2\cos^2\theta}\,
        \delta\theta ,
\label{eq:dphitheta_result}
\end{equation}
where we used Eq.~\eqref{eq:AB}. The recoil $\delta\theta$ is generated by the displacement of Eq.~\eqref{eq:orbital_displacement_operator} and, relative to the emitted state, is of order $\omega/E$. The connection \eqref{eq:dphitheta_result} vanishes as $\beta\to1$ and is largest near threshold.

\subsection{Out-of-plane recoil tilt}
\label{sec:tilt}

The component of the soft recoil odd under $\mathcal R_n$ moves the quark direction out of the production plane, $\delta\hat k=\sin\theta\,\delta\varphi\,\hat n$. Equation~\eqref{eq:delta_X0_theta_collect} shows that this does not generate a rotation about $\hat n$: it produces a $\sigma_n$ component, which lies in the mirror-odd sector and reduces $C_{nn}$. An out-of-plane kick of the quark momentum of size $\delta p_n\sim\omega$ tilts the production plane by $\delta\varphi\simeq\delta p_n/p_T$, where $p_T=E\beta\sin\theta$ is the quark transverse momentum with respect to the photon axis. One therefore expects the mirror-odd kernel to grow as $p_T^{-2}\propto(\beta\sin\theta)^{-2}$. How this tilt is shared between the recoil of $X_0$ itself, whose $\sigma_n$ coefficient in Eq.~\eqref{eq:delta_X0_theta_collect} is $A\cos\theta+B\sin\theta=\sin\theta\cos\theta\,(1-\sqrt{1-\beta^2})$ and vanishes at threshold, the transport of the spin axes and the intrinsic-spin terms depends on the spinor basis, Sec.~\ref{sec:intrinsic}. The coefficient of the pole, which is unity, follows from the complete amplitude in Sec.~\ref{sec:kernels}. A further mirror-odd contribution comes from the incoming-gluon polarization normal to the production plane, which does not contribute at Born level but contributes to $m_0$.

The dominant mirror-odd effect is thus a misalignment, event by event, between the plane of the hard scattering and the production plane reconstructed from the observed momenta. It is nevertheless genuine decoherence of the observed pair: the tilt depends on the momentum and polarization of the unobserved gluon at fixed observed kinematics, so that no choice of axes built from the observed momenta can follow it. The same feature implies that the $p_T^{-2}$ growth signals the breakdown of the expansion as $\omega$ approaches $p_T$, Eq.~\eqref{eq:window}, rather than a physical divergence of the decoherence.

\subsection{Intrinsic-spin contribution}
\label{sec:intrinsic}

The intrinsic-spin part of the next-to-soft operator does not vanish under the tangent projection. Writing $X_0=\boldsymbol\sigma\cdot\mathbf x$ with $\mathbf x=A\hat r+B\hat k$, an intrinsic insertion on the quark side, $(\mathbf a\cdot\boldsymbol\sigma)X_0$, decomposes as
\begin{equation}
(\mathbf a\cdot\boldsymbol\sigma)\,X_0=(\mathbf a\cdot\mathbf x)\,I+i\,(\mathbf a\times\mathbf x)\cdot\hat n\;\sigma_n-i\,a_n\,T_n ,
\label{eq:spin-insertion}
\end{equation}
and an insertion on the antiquark side, $X_0(\mathbf b\cdot\boldsymbol\sigma)$, gives the same structure with $a_n\to-b_n$ in the last term. For the spatial part of the fermion generator, $\epsilon^*_ik_jS^{ij}\propto(\boldsymbol\epsilon^*\times\mathbf k)\cdot\boldsymbol\Sigma$, one has $a_n\propto(\boldsymbol\epsilon^*\times\mathbf k)\cdot\hat n$, which is nonzero for a generic gluon direction and polarization. Its polarization sum is positive,
\begin{equation}
\sum_\lambda\big|(\boldsymbol\epsilon_\lambda^*\times\mathbf k)\cdot\hat n\big|^2=\omega^2\big[1-(\hat\ell\cdot\hat n)^2\big],
\label{eq:spin-polsum}
\end{equation}
so the quadratic intrinsic-spin contribution survives even where its linear average vanishes by symmetry. Thus a spin kick about $\hat n$ feeds the rotation channel, while kicks about in-plane axes feed the mirror-odd channel. Moreover, the split $J_i=L_i+S_i$ depends on the choice of external spin basis: momentum-dependent spinors transfer Wigner-rotation terms, and the transport of the spin axes encountered in Eq.~\eqref{eq:delta_X0_theta_collect}, between the orbital derivative and the explicit spin generator. Only the total next-to-soft amplitude is basis independent. Both $\Gamma_n$ and $\Gamma_\perp$ below are computed from the complete amplitude, including the intrinsic-spin terms of all three legs.

\subsection{Complete next-to-soft kernels}
\label{sec:kernels}

The kernels follow from the complete tree-level amplitude for $\gamma^*_L(q)+g(p)\to q(p_1)+\bar q(p_2)+g(k)$, with open heavy-quark spins, the exact octet color Gram matrix, and the physical polarization sums for both gluons. We expand all eight diagrams of Appendix~B exactly in $\omega$ at fixed observed pair kinematics and obtain $m_0$ in closed form (Appendix~A). Projecting $m_0$ onto the two parity sectors and summing over colors and polarizations gives four angular densities, for $K_\perp$, $K_n$, $K_t$ and $K_a$. They do not depend on $Q^2$ at fixed $(\beta,\theta)$: with $h\equiv Q^2/M^2$ kept as a symbol in the exact expansion, $h$ cancels identically from all four projected densities. In the Born amplitude it enters only through an overall factor. The denominator $p\cdot k$ of the incoming gluon also cancels identically in all of them, so there is no sensitivity to the beam-collinear region. That the projected densities stay finite in this limit follows from collinear factorization. For $k\parallel p$ the amplitude reduces to the $g\to gg$ splitting amplitude times the Born amplitude with incoming momentum $p-k$, at the same observed pair kinematics. The Born amplitude of the longitudinal photon is proportional to $X_0$ for an in-plane polarization of the incoming gluon and vanishes for a normal one, so the collinear-singular part of the amplitude is proportional to $X_0$ and is removed by the projection onto the complement \eqref{eq:complement}. The only remaining angular denominators are the heavy-quark eikonal factors. With $x=\hat\ell\cdot\hat k=\cos\chi$ and after the average over the azimuth of $\hat\ell$ about $\hat k$, each density is a polynomial of degree four in $x$ divided by $(1-\beta^2x^2)^2$. The odd powers integrate to zero over the full solid angle, so that
\begin{equation}
K_X=\frac12\int_{-1}^{1}dx\,\frac{\sum_{j=0}^4n^X_jx^j}{(1-\beta^2x^2)^2}=\sum_{j=0,2,4}n^X_j\,J_j,
\qquad X=\perp,n,t,a,
\label{eq:Kmaster}
\end{equation}
where the coefficients $n^X_j(\beta,\cos\theta)$ are rational functions listed in Appendix~A, Eq.~\eqref{eq:ncoef}. The three integrals
\begin{equation}
J_0=\frac12\left[\frac{1}{1-\beta^2}+L\right],\qquad
J_2=\frac{1}{2\beta^2}\left[\frac{1}{1-\beta^2}-L\right],\qquad
J_4=\frac{1}{2\beta^4}\left[\frac{1}{1-\beta^2}+2-3L\right],
\qquad
L(\beta)=\frac{1}{2\beta}\ln\frac{1+\beta}{1-\beta},
\label{eq:Lbeta}
\end{equation}
contain the single transcendental function $L(\beta)$. Collecting terms gives, with $c\equiv\cos\theta$, $s\equiv\sin\theta$ and $D\equiv1-\beta^2c^2$, the closed forms
\begin{align}
K_\perp(\beta,\theta)&=\frac{P_\perp\,L-Q_\perp}{24\,\beta^4s^2D},
&
K_n(\beta,\theta)&=\frac{(1-\beta^2)\,P_n\,L-Q_n}{24\,\beta^2D^2},
\label{eq:Kclosed}\\
K_t(\beta,\theta)&=-\frac{P_t\,L+Q_t}{12\,\beta^4s^2D},
&
K_a(\beta,\theta)&=\frac{\sqrt{1-\beta^2}\,\big[Q_a-P_a\,L\big]}{12\,\beta^3sD},
\label{eq:Kta}
\end{align}
where $t=(\alpha_s/2\pi)\tau_c^2K_t$ and $a=(\alpha_s/2\pi)\tau_c^2K_a$ are the trace and antisymmetric in-plane coefficients of Eq.~\eqref{eq:Csoft}, and the polynomials are
\begin{align}
P_\perp&=\beta^6c^2(1+c^2)+\beta^4(1-59c^2+24c^4)+\beta^2(48-25c^2+39c^4)+15-45c^2,\nonumber\\
Q_\perp&=\beta^4c^2(37c^2-35)+\beta^2(29-40c^2+39c^4)+15-45c^2,\nonumber\\
P_n&=\beta^4c^2(c^2-3)+\beta^2(3+80c^2-51c^4)+15-45c^2,\nonumber\\
Q_n&=\beta^4c^2(35c^2-33)-\beta^2(31-110c^2+51c^4)+15-45c^2,\nonumber\\
P_t&=\beta^6c^2(3c^2-5)+\beta^4(7+107c^2-80c^4)+\beta^2(72-275c^2+141c^4)+45c^2-15,\nonumber\\
Q_t&=\beta^4c^2(33c^2-31)-\beta^2(91-260c^2+141c^4)+15-45c^2,\nonumber\\
P_a&=30+2\beta^2-45c^2+14\beta^2c^2-\beta^4c^2,\qquad Q_a=30-45c^2-\beta^2c^2 .
\label{eq:Kpolys}
\end{align}
Near threshold,
\begin{equation}
K_\perp\simeq\frac{1}{\beta^2\sin^2\theta}+\frac{15-13\cos^2\theta}{18\sin^2\theta},
\qquad
K_n\simeq1,
\qquad \beta\to0 .
\label{eq:Kthreshold}
\end{equation}
The pole in $p_T=E\beta\sin\theta$ reflects the out-of-plane tilt of Sec.~\ref{sec:tilt}; its coefficient is unity. It implies that the expansion parameter of the mirror-odd channel is $E_{\rm cut}/p_T$ rather than $E_{\rm cut}/E$, so that the present fixed-order result requires
\begin{equation}
\Lambda_{\rm QCD}\ll E_{\rm cut}\ll p_T=E\beta\sin\theta ,
\label{eq:window}
\end{equation}
which excludes the forward and backward regions and the threshold region. In the opposite limit, $\beta\to1$ at fixed $\theta$, one has $Q_n\to-16\sin^4\theta$, $P_\perp\to64\sin^4\theta$ and $Q_\perp\to-4\sin^2\theta\,(19\cos^2\theta-11)$, so that
\begin{equation}
K_n\to\frac23,
\qquad
K_\perp\simeq\frac83\,L(\beta)+\frac{19\cos^2\theta-11}{6\sin^2\theta},
\qquad \beta\to1 .
\label{eq:Kultrarel}
\end{equation}
The ratio $K_\perp/K_n$ therefore diverges as $(\beta\sin\theta)^{-2}$ near threshold, Eq.~\eqref{eq:Kthreshold}, and logarithmically, as $4L(\beta)$, as $\beta\to1$. In between we have no analytic bound; evaluating Eqs.~\eqref{eq:Kclosed}--\eqref{eq:Kpolys} on a $200\times200$ grid covering $0.02\le\beta\le0.99$ and $0<\theta<\pi$ gives $K_\perp/K_n\ge3.5$, so that $K_\perp>2K_n$ on the whole grid. For $\beta\lesssim0.05$ the closed forms involve cancellations of order $\beta^{-4}$, and Eq.~\eqref{eq:Kthreshold} or the representation \eqref{eq:Kmaster} should be used instead.

The measurement of Sec.~\ref{sec:eic_observable} excludes cones of half-opening angle $R_B$ about the target direction and $R$ about the heavy-quark directions. The corresponding wide-angle kernels are $K^{\rm wa}=K-\Delta K(R,R_B)$, where $\Delta K$ is the integral of the same densities over the excluded cones. The target cone breaks the azimuthal symmetry about $\hat k$, and for $R=R_B=0.5$ we evaluate $\Delta K$ numerically. It is a smooth correction: for $0.2\le\beta\le0.9$ and $\pi/6\le\theta\le\pi/2$ it removes $19$--$30\%$ of $K_\perp$ and $11$--$50\%$ of $K_n$. It grows toward $\beta\to1$: for the $c\bar c$ benchmark of Table~\ref{tab:bench} ($\beta=0.93$, $\theta=\pi/2$) the full-angle values $K_\perp=3.46$ and $K_n=0.77$ become $K^{\rm wa}_\perp=2.27$ and $K^{\rm wa}_n=0.32$, a reduction by $34\%$ and $59\%$. The wide-angle inequality $K^{\rm wa}_\perp>2K^{\rm wa}_n$ has been checked in this range and at all benchmarks of Table~\ref{tab:bench}, where $K^{\rm wa}_\perp/K^{\rm wa}_n=3.7$--$7.1$. The benchmarks of Sec.~\ref{sec:eic_observable} use $K^{\rm wa}$.

\begin{figure}[t]
    \centering
    \includegraphics[width=0.98\linewidth]{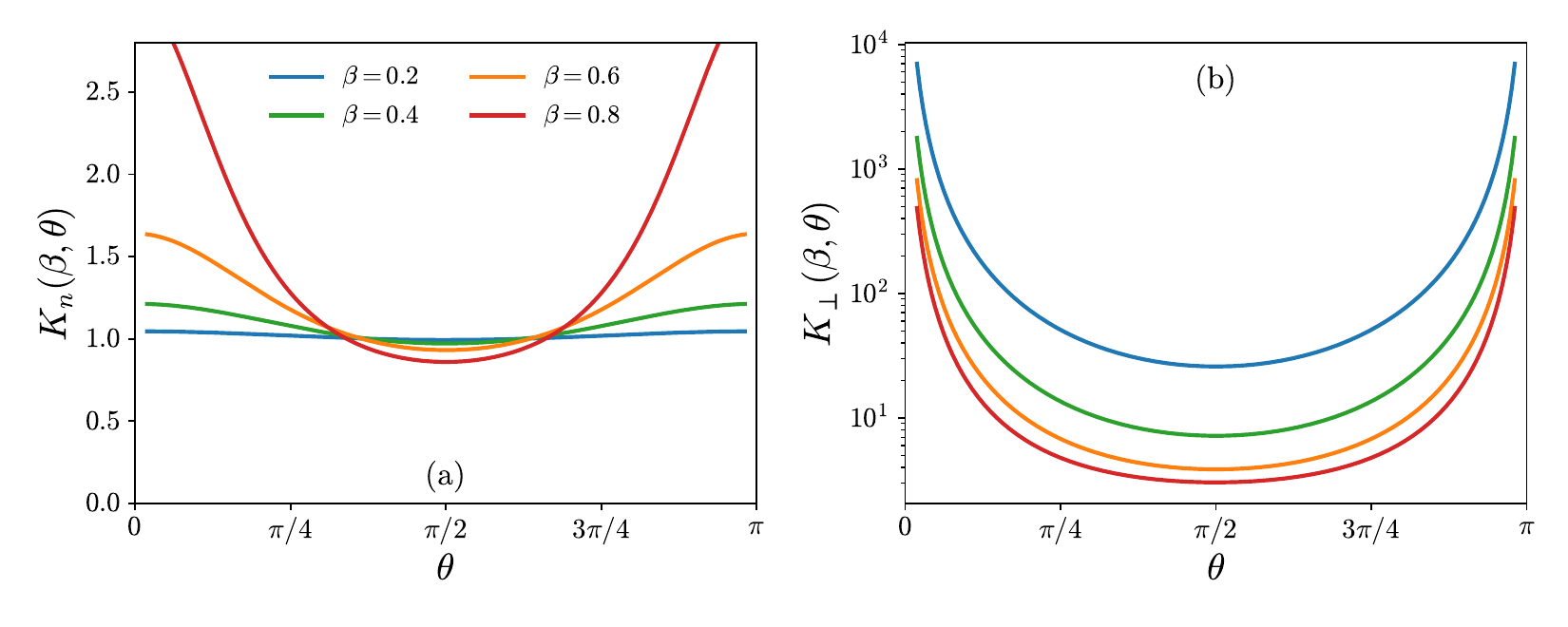}
\caption{Next-to-soft kernels of the mirror-even rotation channel, $K_n$ (a), and of the mirror-odd normal-spin-flip channel, $K_\perp$ (b), as functions of the heavy-quark scattering angle $\theta$ in the $q\bar q$ center-of-mass frame, for $\beta=0.2,0.4,0.6,0.8$. The curves are the closed forms \eqref{eq:Kclosed} for the full solid angle. The dephasing coefficients are $\Gamma_{n,\perp}=(\alpha_s/\pi)\tau_c^2K_{n,\perp}$ for a cumulative resolution $\tau_c=E_{\rm cut}/E$, Eq.~\eqref{eq:Gammas}. The rise of $K_\perp$ toward $\theta\to0,\pi$ follows $(\beta\sin\theta)^{-2}$, Eq.~\eqref{eq:Kthreshold}, and marks the region in which the expansion in $E_{\rm cut}/p_T$ fails. For a given resolution the curves are predictive only where $\tau_c\lesssim\beta\sin\theta/2$, i.e.\ $E_{\rm cut}\lesssim p_T/2$; for $\tau_c=0.2$ this excludes $\sin\theta<0.4/\beta$, and the whole angular range for $\beta\le0.4$.}
    \label{fig:kernels}
\end{figure}

\section{Decoherence channels: stochastic rotation and normal-spin flip}
\label{sec:dephasing}

The decohering map \eqref{eq:drho} is block diagonal in the two parity sectors of Eq.~\eqref{eq:blockdiag}. In this section we write the two blocks in operator form and combine them into the corrected spin-correlation matrix.

\subsection{Mirror-even channel: rotation of the spin-correlation plane}

For each fixed unresolved gluon $\xi$, the mirror-even displacement is a coherent rotation of the two spins about $\hat n$,
\begin{equation}
U(\xi)=\exp\left[-\frac{i}{2}\delta\phi_n(\xi)J_n\right],
\qquad
J_n=\sig_n\otimes I+I\otimes\sig_n .
\label{eq:U}
\end{equation}
A more general treatment introduces two angles, $J_n\delta\phi_n\to\delta\phi_q\,\sig_n\otimes I+\delta\phi_{\bar q}\,I\otimes\sig_n$, under which the in-plane correlation block transforms as $C_\parallel\to R(\delta\phi_q)\,C_\parallel\,R^{T}(\delta\phi_{\bar q})$. Decomposing a general block into its scalar part $(C_{rr}+C_{kk})/2$, its antisymmetric part $a=(C_{kr}-C_{rk})/2$, and its symmetric-traceless part ${\cal Q}=(C_{rr}-C_{kk})/2+i(C_{rk}+C_{kr})/2$, one finds that the first two rotate with the difference $\delta\phi_q-\delta\phi_{\bar q}$ (and are invariant under a common rotation), while ${\cal Q}\to e^{\pm i(\delta\phi_q+\delta\phi_{\bar q})}{\cal Q}$ carries rotation charge two. The longitudinal Born block of Eq.~\eqref{eq:clo} has $C_{rr}=-C_{kk}$ and $C_{rk}=C_{kr}$, so it is purely symmetric traceless and depends only on the sum $\delta\phi_q+\delta\phi_{\bar q}$. For this state the common rotation \eqref{eq:U} with $\delta\phi_n=(\delta\phi_q+\delta\phi_{\bar q})/2$ is therefore exact at this order rather than an approximation. Moreover, the mirror-even complement of a pure maximally entangled state is one dimensional (along $T_n$), so a second angle carries no additional information at this order. For fixed $\xi$,
\begin{equation}
\rho(\xi)=U\rho^{(0)}U^\dagger=\rho^{(0)}-\frac{i}{2}\delta\phi_n[J_n,\rho^{(0)}]
-\frac18\delta\phi_n^2[J_n,[J_n,\rho^{(0)}]]+\calO(\delta\phi_n^3),
\label{eq:rho-expand}
\end{equation}
which is the Baker-Campbell-Hausdorff expansion of a unitary conjugation. The in-plane recoil $\delta\theta$ is even under $\mathcal R_n$, so the inclusive mean $\langle\delta\phi_n\rangle$ need not vanish. The linear term is a coherent rotation of the spin-correlation plane by an angle of order $\alpha_s\tau_c$; it rotates $C_{rr}$ and $C_{rk}$ into each other but does not change $|{\cal Q}|$, the purity or the entanglement. The quadratic term, averaged with the inclusive measure of Eq.~\eqref{eq:Sigma}, is the dephasing map of the rotation channel,
\begin{equation}
\delta\rho_+=-\frac18\langle\delta\phi_n^2\rangle\,[J_n,[J_n,\rho^{(0)}]]
=\frac{\Gamma_n}{8}\left[J_n\rho^{(0)} J_n-\frac12\{J_n^2,\rho^{(0)}\}\right],
\qquad \Gamma_n=2\langle\delta\phi_n^2\rangle .
\label{eq:rhoav}
\end{equation}
It is trace and Hermiticity preserving, and $L_n=\sqrt{\gamma_n}\,J_n$ with $\gamma_n=\Gamma_n/8=\langle\delta\phi_n^2\rangle/4$ is its Lindblad jump operator. Since $[J_n,[J_n,{\cal O}]]=16\,{\cal O}$ for any charge-two operator, Eq.~\eqref{eq:rhoav} multiplies ${\cal Q}$ by $1-2\langle\delta\phi_n^2\rangle=1-\Gamma_n$. Because $[J_n,\Pi_n]=0$, it leaves $C_{nn}$ unchanged.

\subsection{Mirror-odd channel: normal-spin flip}

Because the Born state is maximally entangled, every state in its $\Pi_n=-1$ complement is obtained by a single-spin operation about an in-plane axis: $\sigma_rX_0=A\,I-iB\,\sigma_n$ and $\sigma_kX_0=B\,I+iA\,\sigma_n$, by Eq.~\eqref{eq:pauli-triad}, span $\{I,\sigma_n\}$. The mirror-odd part of Eq.~\eqref{eq:drho} therefore acts on the Born state as a normal-spin-flip channel,
\begin{equation}
\delta\rho_-=\sum_{a,b\in\{r,k\}}\gamma^{\perp}_{ab}\left[\sigma^{(q)}_a\rho^{(0)}\sigma^{(q)}_b-\delta_{ab}\,\rho^{(0)}\right],
\qquad \sigma^{(q)}_a=\sigma_a\otimes I,
\qquad {\rm Tr}\,\gamma^\perp=\frac{\Gamma_\perp}{2},
\label{eq:rho-odd}
\end{equation}
with $\gamma^\perp$ a positive $2\times2$ matrix obtained from $\Sigma_{AB}$ in the $\{I,\sigma_n\}$ block. Equivalently, the flips may be attributed to the antiquark. Since $\{\sigma^{(q)}_a,\Pi_n\}=0$ for $a=r,k$, each flip reverses the normal-spin parity.

\subsection{Soft-recoil corrected spin-correlation matrix}

Combining Eqs.~\eqref{eq:rhoav} and \eqref{eq:rho-odd}, the normalized soft-recoil corrected longitudinal spin-correlation matrix at fixed order is
\begin{equation}
C^{L}_{\rm soft}
=
\begin{pmatrix}
1-\Gamma_\perp & 0 & 0\\
0 & \lambda\,C_{rr}+\frac{t}{2} & \lambda\,C_{rk}-a\\
0 & \lambda\,C_{kr}+a & \lambda\,C_{kk}+\frac{t}{2}
\end{pmatrix},
\qquad
\lambda=1-\Gamma_n-\frac{\Gamma_\perp}{2},
\qquad
|t+2ia|\le\Gamma_\perp ,
\label{eq:Csoft}
\end{equation}
up to the coherent in-plane rotation discussed above, which acts only on the symmetric-traceless part. The single-spin polarizations remain zero. Here $t=C_{rr}+C_{kk}$ and $a$ are generated by the mirror-odd channel alone, $t=(\alpha_s/2\pi)\tau_c^2K_t$ and $a=(\alpha_s/2\pi)\tau_c^2K_a$ with $K_{t,a}$ given in Eq.~\eqref{eq:Kta}; the bound $|t+2ia|\le\Gamma_\perp$ follows from the positivity of $\gamma^\perp$. Positivity holds by construction, since $\Sigma_{AB}$ is a Gram matrix contracted with a positive-definite color matrix, and the bound is satisfied by Eqs.~\eqref{eq:Kclosed}--\eqref{eq:Kpolys} throughout the physical range. In Eq.~\eqref{eq:Csoft} the normal correlation is reduced only by the mirror-odd channel. The rotation-invariant in-plane magnitude
\begin{equation}
|{\cal Q}|=\frac12\sqrt{(C_{rr}-C_{kk})^2+(C_{rk}+C_{kr})^2}\;\longrightarrow\;1-\Gamma_n-\frac{\Gamma_\perp}{2}
\label{eq:Qmag}
\end{equation}
is reduced by both channels. A suppression of $C_{rr}$, $C_{kk}$ and $C_{rk}$ by one common factor holds only for the symmetric-traceless part, and it relies on the special Born relations $C_{rr}=-C_{kk}$ and $C_{rk}=C_{kr}$. The trace $C_{rr}+C_{kk}$ has rotation charge zero, is invariant under any common rotation about $\hat n$, and is generated here only by the mirror-odd channel. Individual entries also receive the coherent ${\cal O}(\alpha_s\tau_c)$ rotation. For these reasons the observables of Sec.~\ref{sec:eic_observable} are built from $C_{nn}$ and $|{\cal Q}|$. They are also insensitive to the interference terms $m_{-1}m_0^\dagger$ and $m_{-1}m_1^\dagger$, which are linear in the displacement and enter the unnormalized density matrix as $|B\rangle\langle D|+|D\rangle\langle B|$, with $|B\rangle$ the Born state. For $C_{nn}$ this is exact: $C_{nn}={\rm Tr}[\rho\,\Pi_n]$ is linear in $\rho$, and $\langle D|\Pi_n|B\rangle=\langle D|B\rangle$ vanishes for any $|D\rangle$ orthogonal to the Born state, while a component along $|B\rangle$ only changes the normalization and enters $1-C_{nn}$ at higher order. For $|{\cal Q}|$ we use that $|{\cal Q}|$ is bounded by the largest singular value of the in-plane block, which is at most one, and that the Born state saturates the bound. A linear deformation therefore leaves $|{\cal Q}|$ unchanged at first order and contributes only through its square. For $m_{-1}m_1^\dagger$ this is ${\cal O}(\tau^4)$. For $m_{-1}m_0^\dagger$ it is the square of the mean displacement: negligible in the inclusive state, ${\cal O}(\alpha_s^2\tau^2)$, and evaluated for the conditional state in Sec.~\ref{sec:resolution}. To the order considered, $C_{nn}$ and $|{\cal Q}|$ are therefore determined by the Gram matrix of $m_0$.

For the cumulative resolution $\tau_c$, the coefficients are
\begin{equation}
\Gamma_n(\beta,\theta)=\frac{\alpha_s}{\pi}\,\tau_c^2\,K_n(\beta,\theta),
\qquad
\Gamma_\perp(\beta,\theta)=\frac{\alpha_s}{\pi}\,\tau_c^2\,K_\perp(\beta,\theta),
\qquad
\tau_c=\frac{E_{\rm cut}}{E},
\label{eq:Gamma}
\end{equation}
with $K_{n,\perp}$ given in closed form by Eq.~\eqref{eq:Kclosed} for the full solid angle and replaced by $K^{\rm wa}_{n,\perp}$ for the wide-angle region (Sec.~\ref{sec:kernels}). Since $K_\perp>2K_n$, the normal correlation is degraded more strongly than the in-plane coherence, which is the opposite of the anisotropy of a pure rotation channel.

\subsection{Multiple soft emissions and exponentiation}
\label{sec:multiple}

The result \eqref{eq:Csoft} was obtained from a single unresolved emission. For several emissions $k_1,\ldots,k_N$ in the rotation channel the accumulated angle is $\Delta\phi_n=\sum_a\delta\phi_n(k_a)$. Under a common rotation by $\Delta\phi_n$ the charge-two tensor transforms as ${\cal Q}\to e^{2i\Delta\phi_n}{\cal Q}$, so the in-plane coherence is governed by the charge-two characteristic function
\begin{equation}
\left\langle e^{2i\Delta\phi_n}\right\rangle=\exp\left[\sum_{m\ge1}\frac{(2i)^m}{m!}\,\kappa_m\right],
\qquad \kappa_m=\langle\Delta\phi_n^m\rangle_c .
\label{eq:charfun}
\end{equation}
Odd cumulants produce phases, that is, coherent rotations, while the magnitude is $\exp[-2\kappa_2+\frac{2}{3}\kappa_4-\cdots]$. Reflection symmetry alone does not make the distribution Gaussian; the higher even cumulants are separate dynamical quantities. If the emissions form an independent Poisson process with intensity $d\nu(k)$, then $\ln\langle e^{2i\Delta\phi_n}\rangle=\int d\nu\,[e^{2i\delta\phi_n(k)}-1]$ exactly, so $\kappa_m=\int d\nu\,\delta\phi_n^m$ and $\kappa_4/\kappa_2=O(\omega^2/E^2)$. In this case, and only in this case, the Gaussian form follows from the smallness of the single-emission kick. The accumulated variance $\kappa_2$ is then the sum of the single-emission variances, and at fixed order $\kappa_2=\langle\delta\phi_n^2\rangle=\Gamma_n/2$. Hence
\begin{equation}
 \left\langle e^{2i\Delta\phi_n}\right\rangle
 =\exp\left[-2\left\langle\Delta\phi_n^2\right\rangle\right]
 =\exp[-\Gamma_n] ,
 \label{eq:exp-n}
\end{equation}
which expands to the fixed-order factor $1-\Gamma_n$ of Eq.~\eqref{eq:Csoft}. In the mirror-odd channel each emission reverses $\Pi_n$ with a small probability; for independent emissions $C_{nn}=\langle(-1)^{N_{\rm flip}}\rangle=\exp(-\Gamma_\perp)$. Within the same model,
\begin{equation}
 C_{\rm soft}^{L,\,{\rm exp}}:\qquad C_{nn}=e^{-\Gamma_\perp},\qquad |{\cal Q}|\simeq e^{-\Gamma_n-\Gamma_\perp/2}.
 \label{eq:Cexp}
\end{equation}
Equations~\eqref{eq:exp-n}--\eqref{eq:Cexp} are a model for the accumulation of independent emissions, not a resummation derived from QCD, and the fixed-order map \eqref{eq:Csoft} remains the primary result. The kernels contain no large logarithm of the resolution, so the decoherence coefficients themselves require no resummation; the probability of emission into a finite energy shell can nevertheless be of order one, and its effect on the radiation bins is estimated in Sec.~\ref{sec:multi-bins}. Within the window \eqref{eq:window} the coefficients are small, $\Gamma_{n,\perp}\lesssim\text{few}\times10^{-2}$, so that the exponentiated and fixed-order forms differ by less than $10^{-3}$. Correlated multi-soft emission, non-global configurations, and color evolution can violate the independence assumption, on which Eq.~\eqref{eq:Cexp} depends and the fixed-order map \eqref{eq:Csoft} does not.

\section{Estimates of Entanglement Measures and Bell Violation}
\label{sec:measures}

We now estimate the degradation of entanglement and Bell nonlocality induced by the two channels. By Eq.~\eqref{eq:blockdiag}, the reduced state is block diagonal in $\Pi_n$, and to first order in $\Gamma_{n,\perp}$ its eigenvalues are
\begin{equation}
 p_1=1-\frac{\Gamma_n+\Gamma_\perp}{2},
 \qquad
 p_2=\frac{\Gamma_n}{2},
 \qquad
 p_3+p_4=\frac{\Gamma_\perp}{2},
 \label{eq:bell_eigenvalues_exp}
\end{equation}
where $p_2$ belongs to the rotated in-plane state and $p_{3,4}$ are the eigenvalues of $\gamma^\perp$ in the mirror-odd sector. At leading order $\Gamma_n=\Gamma_\perp=0$, so that $p_1=1$, as expected for a pure maximally entangled state. In the rotation-only limit $\Gamma_\perp=0$ the state is the one-parameter dephased Bell state with $p_{1,2}=(1\pm\lambda_n)/2$ and $\lambda_n=1-\Gamma_n$ (or $e^{-\Gamma_n}$ after exponentiation). The formulas below reduce to that case. In general the reduced state also depends on $t$ and $a$, which split $p_3$ and $p_4$; to first order, however, the purity, the concurrence and the negativity depend only on $\Gamma_n+\Gamma_\perp$, and the Bell parameter only on the singular values given below.

\subsubsection{Purity}

The purity of a two-spin state with vanishing single-particle polarizations is ${\cal P}={\rm Tr}\,\rho^2=\frac14\big(1+\sum_{ij}C_{ij}^2\big)$. Using Eq.~\eqref{eq:Csoft},
\begin{equation}
 {\cal P}
 =
 1-\Gamma_n-\Gamma_\perp+\mathcal{O}(\Gamma^2).
 \label{eq:purity_exp}
\end{equation}
In the rotation-only limit this reproduces ${\cal P}=(1+\lambda_n^2)/2$.

\subsubsection{Linear Entropy}

The linear entropy $S_L=1-{\rm Tr}\,\rho^2$ is
\begin{equation}
 S_{L}
 =
 \Gamma_n+\Gamma_\perp+\mathcal{O}(\Gamma^2).
 \label{eq:linear_entropy_exp}
\end{equation}
It is fixed by the trace of the Gram matrix \eqref{eq:Sigma} and is independent of the decomposition into channels.

\subsubsection{Von Neumann Entropy}

The von Neumann entropy $S_{\rm vN}=-\sum_ip_i\ln p_i$ follows from Eq.~\eqref{eq:bell_eigenvalues_exp}. For small $\Gamma_{n,\perp}$ it is dominated by the small eigenvalues,
\begin{equation}
 S_{\rm vN}\simeq\frac{\Gamma_n}{2}\left(1-\ln\frac{\Gamma_n}{2}\right)-\sum_{i=3,4}p_i\ln p_i+\frac{\Gamma_\perp}{2},
 \label{eq:vn_entropy_exp}
\end{equation}
which vanishes when $\Gamma_n=\Gamma_\perp=0$ and grows as unresolved soft radiation turns the initially pure Bell-like state into a mixed state. Unlike $S_L$, it is not fixed by $\Gamma_n+\Gamma_\perp$ alone, since it depends on how $\Gamma_\perp/2$ is split between $p_3$ and $p_4$ by $t$ and $a$.

\subsubsection{Concurrence}

Since the single-spin polarizations vanish, the state of Eq.~\eqref{eq:Csoft} is related by local rotations to a state diagonal in a Bell basis, with eigenvalues \eqref{eq:bell_eigenvalues_exp}. Its concurrence is then $\max(0,2p_{1}-1)$, so that
\begin{equation}
 {\cal C}=1-\Gamma_n-\Gamma_\perp+\mathcal{O}(\Gamma^2) .
 \label{eq:concurrence_exp}
\end{equation}
In the rotation-only limit it reduces to ${\cal C}=\lambda_n$. The independent-emission form is ${\cal C}\simeq e^{-\Gamma_n-\Gamma_\perp}$.

\subsubsection{Negativity}

The negativity of the same state is
\begin{equation}
 {\cal N}
 =
 \frac{1}{2}\left(1-\Gamma_n-\Gamma_\perp\right)+\mathcal{O}(\Gamma^2),
 \qquad
 \Delta{\cal N}={\cal N}_{\rm LO}-{\cal N}=\frac{\Gamma_n+\Gamma_\perp}{2} ,
 \label{eq:negativity_exp}
\end{equation}
with ${\cal N}_{\rm LO}=1/2$.

\subsubsection{Bell-CHSH Violation}

The maximal Bell-CHSH parameter is determined by the two largest singular values of the correlation matrix, $B_{\rm max}=2\sqrt{s_{(1)}^2+s_{(2)}^2}$. Equation~\eqref{eq:Csoft} is block diagonal, with $s_{nn}=|C_{nn}|$. Its in-plane block is $C_\parallel=\lambda R+P$, where $R$ is the Born block, a reflection matrix ($R^T=R$, $R^2=I$), and $P=\frac t2I+aJ$ with $J$ the $2\times2$ rotation generator of Appendix~A. Since a reflection anticommutes with $J$, $C_\parallel^TC_\parallel=\lambda^2I+\lambda R(t+2aJ)+{\cal O}(\Gamma^2)$, and $R(t+2aJ)$ is traceless with determinant $-(t^2+4a^2)$, hence has eigenvalues $\pm|t+2ia|$. Retaining $t$ and $a$ throughout, the singular values are therefore
\begin{equation}
 s_{nn}=1-\Gamma_\perp,
 \qquad
 s_\pm=1-\Gamma_n-\frac{\Gamma_\perp}{2}\pm\frac{|t+2ia|}{2},
 \label{eq:singular}
\end{equation}
so that, to first order,
\begin{equation}
 B_{\rm max}=2\sqrt2\left[1-\frac{\delta_{(1)}+\delta_{(2)}}{2}\right],
 \label{eq:bell_exp}
\end{equation}
where $\delta_{(1)},\delta_{(2)}$ are the two smallest of the deficits $1-s_{nn}$ and $1-s_\pm$. In the rotation-only limit, $\Gamma_\perp=0$, this reduces to $B_{\rm max}=2\sqrt{1+\lambda_n^2}$. 

The entanglement and the purity are thus fixed by $\Gamma_n+\Gamma_\perp=2\,{\rm Tr}\,\Sigma$, and the Bell parameter by the two smallest deficits. For the inclusive state with $\alpha_s=0.25$ and $\tau_c=0.2$, $\Gamma_n+\Gamma_\perp$ is of order $10^{-2}$ in the central region; in the radiation-binned samples of Sec.~\ref{sec:eic_observable} it reaches a few per cent.

\section{EIC observable: radiation-binned spin-dephasing ratio}
\label{sec:eic_observable}

The soft-recoil mechanism discussed here leads to a distinctive experimental signature: unresolved radiation degrades the normal spin correlation and the in-plane spin coherence by different, calculable amounts, Eq.~\eqref{eq:Csoft}. We therefore compare the spin correlations in bins of the radiation defined below. \\

We consider semi-inclusive heavy-flavor production in deep-inelastic
scattering,
\begin{equation}
    e(\ell) + p(P) \to e(\ell') + q(p_1) + \bar q(p_2) + X ,
\end{equation}
where \(q=\ell-\ell'\) is the virtual-photon momentum,
\(Q^2=-q^2\), and the heavy quark and antiquark are reconstructed
through heavy-flavour jets or through identified heavy-flavor hadrons, for example $e p \to e' + J_q + J_{\bar q} + X$ or, in more spin-sensitive channels, $e p \to e' + \Lambda_q + \bar\Lambda_q + X$ with $q=c,b$. The latter class is useful because the spin of the heavy quark is partially transferred to the heavy baryon and analyzed through its decay products. For each event the Born variables $\Phi_B$ are reconstructed from the scattered lepton and the two heavy-flavor axes, and the production plane is defined by these axes and the photon direction in the pair rest frame. If \(a\) and \(b\) denote the spin analyzers associated with the heavy quark and antiquark, their angular distribution is
\begin{equation}
    \frac{1}{\sigma} \frac{d\sigma}{d\Omega_a d\Omega_b} =
    \frac{1}{(4\pi)^2} \left[ 1 + \kappa_a \kappa_b \sum_{i,j=r,n,k} C_{ij}\, \hat a_i \hat b_j \right],
    \qquad \kappa_{a}=\alpha_{a}\,{\cal T}_{a},
    \label{eq:spin_analyzer_distribution}
\end{equation}
where \(\hat a_i\) and \(\hat b_j\) are the analyzer directions in the event basis, \(\alpha_{a,b}\) are the weak-decay analyzing powers, and ${\cal T}_{a,b}$ are the spin-transfer coefficients from the heavy quark to the reconstructed hadron. Below we write $\kappa\equiv\kappa_a\kappa_b$ for the total dilution.

\subsection{Infrared-safe radiation resolution and radiation bins}
\label{sec:resolution}

To connect with unresolved soft radiation we define the extra-radiation variable as a wide-angle energy flow in the Born-projected pair rest frame,
\begin{equation}
    E_{\rm rad} = \sum_{h\in X} (u_B\cdot p_h)\,\Theta_{\rm wa}(\hat p_h) ,
    \label{eq:erad_def}
\end{equation}
where $u_B^\mu$ is the four-velocity of the Born-projected heavy pair, and $\Theta_{\rm wa}$ restricts the sum to the region $\Omega_{\rm wa}$ outside cones of half-opening angle $R_B$ about the target direction and $R$ about the two heavy-flavor axes. The sum runs over hadrons other than the reconstructed heavy hadrons, their decay products, and the soft companion pions of $\Sigma_{c,b}^{(*)}\to\Lambda_{c,b}\,\pi$ feed-down. These must be removed explicitly: in the pair rest frame $E=4$--$8$~GeV, so the heavy hadrons have $\gamma\beta\sim1$--$2$, and for example the daughters of $\Lambda_c^+\to\Lambda\pi^+$, with a decay momentum of $0.86$~GeV, frequently leave cones of radius $R=0.5$. The axes are obtained with an infrared- and collinear-safe flavor-aware jet algorithm, such as flavor-$k_T$ or flavored anti-$k_T$ \cite{Banfi:2006hf,Czakon:2022wam}. $E_{\rm rad}$ is soft safe, since a soft particle of momentum $\lambda p$ changes it by ${\cal O}(\lambda)$. It is collinear safe, since away from the region boundaries a splitting $p\to zp+(1-z)p$ leaves $u_B\cdot p$ unchanged. The beam cone removes the target remnant and the initial-state collinear region, whose singular part belongs to the gluon distribution. The heavy-flavor cones remove the final-state quasi-collinear region, discussed in Sec.~\ref{sec:collinear}. The observable is non-global. With a single emission this has no effect; non-global logarithms of the resolution first arise at ${\cal O}(\alpha_s^2)$ and are not included here. For a single soft gluon $E_{\rm rad}=\omega\,\Theta_{\rm wa}(\hat\ell)$, so the measurement reduces to the energy bounds below inside $\Omega_{\rm wa}$, which is the region used for the kernels $K^{\rm wa}$.

$E_{\rm rad}$ also receives non-perturbative contributions. The Born pair is a color octet whose color is neutralized by the target remnant, so that hadronization populates the wide-angle region between the pair and the remnant even in the absence of a perturbative emission; in addition, each hadron contributes at least its mass to $u_B\cdot p_h$. If these effects are modeled by a shift $E_{\rm rad}\to E_{\rm rad}+\delta_{\rm NP}$ of the perturbative distribution, a bin with edges $E^{(i)}_{\rm cut}$ selects perturbative configurations with edges $E^{(i)}_{\rm cut}-\delta_{\rm NP}$, and since the coefficients scale as $\tau^2$, the low-bin coefficient changes by the factor $(1-\delta_{\rm NP}/E^{(1)}_{\rm cut})^2$. With $\delta_{\rm NP}$ of a few hundred MeV this is not a small correction at $E^{(1)}_{\rm cut}=0.4$--$0.8$~GeV. A mass-insensitive variant, with $u_B\cdot p_h$ replaced by the three-momentum $|\mathbf p_h|$ in the pair rest frame, removes the hadron-mass part of $\delta_{\rm NP}$ but not the hadronization of the octet color flow, which has to be determined from data or from a hadronization model in each radiation bin.

\begin{figure}[t]
    \centering
    \includegraphics[width=0.98\linewidth]{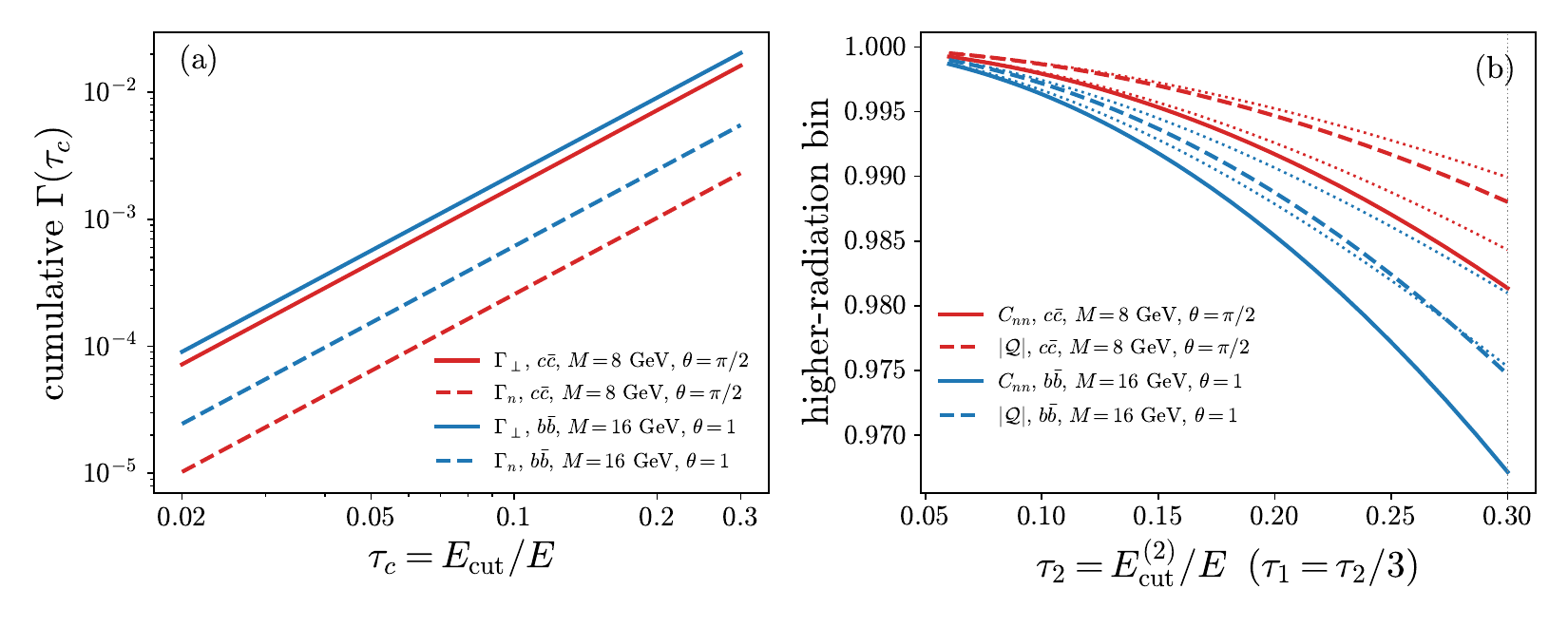}
\caption{Resolution dependence of the soft-recoil decoherence for a $c\bar c$ pair with $M=8$~GeV ($\beta=0.93$) at $\theta=\pi/2$ and a $b\bar b$ pair with $M=16$~GeV ($\beta=0.80$) at $\theta=1$. (a) Cumulative coefficients $\Gamma_\perp$ (solid) and $\Gamma_n$ (dashed) of the inclusive state, Eq.~\eqref{eq:Gamma} with $K\to K^{\rm wa}$, for $\alpha_s=0.25$, showing the $\tau_c^2$ scaling. (b) Normal correlation $C_{nn}$ (solid) and in-plane magnitude $|{\cal Q}|$ (dashed) in the higher-radiation bin $\tau_1\le E_{\rm rad}/E<\tau_2$ with $\tau_1=\tau_2/3$, Eq.~\eqref{eq:Gamma-high}; the dotted lines include the calculable next-to-soft correction to the emission rate in the normalization (Sec.~\ref{sec:resolution}). The vertical line marks the upper end, $\tau_2=0.3$, of the range used. Wide-angle region with cones $R=R_B=0.5$. The curves in (b) are leading-power, single-emission results with an ${\cal O}(\tau_2)$ uncertainty of about $30\%$. At fixed $\tau_2/\tau_1$ the independent-emission estimate of Sec.~\ref{sec:multi-bins} rescales $1-C_{nn}$ and $1-|{\cal Q}|$ in this bin by a $\tau_2$-independent factor, $0.66$ for the $c\bar c$ and $0.73$ for the $b\bar b$ configuration.}
    \label{fig:resolution}
\end{figure}

We define two non-overlapping radiation bins,
\begin{align}
    \text{low-radiation bin:}
    \qquad
    & 0 \le E_{\rm rad} < E_{\rm cut}^{(1)}, \nonumber
    \\
    \text{higher-radiation bin:}
    \qquad
    & E_{\rm cut}^{(1)} \le E_{\rm rad} < E_{\rm cut}^{(2)},
    \qquad \tau_{1,2}\equiv\frac{E_{\rm cut}^{(1,2)}}{E},
    \label{eq:bins}
\end{align}
with $E_{\rm cut}^{(2)}$ inside the window \eqref{eq:window}. The two bins have different integration limits and different normalizations. The low-radiation bin contains the Born, the virtual correction, and real emission with $\omega<E_{\rm cut}^{(1)}$. Its state is the inclusive map \eqref{eq:drho} with cumulative resolution $\tau_1$:
\begin{equation}
\Gamma_{n,\perp}^{\rm low}=\frac{\alpha_s}{\pi}\,\tau_1^2\,K^{\rm wa}_{n,\perp}(\beta,\theta).
\label{eq:Gamma-low}
\end{equation}
At ${\cal O}(\alpha_s)$ the higher-radiation bin contains only events with one wide-angle emission in the shell $E_{\rm cut}^{(1)}\le\omega<E_{\rm cut}^{(2)}$, whose energy is resolved but whose direction, polarization and color are not. Its state is the conditional state of Eqs.~\eqref{eq:cond-cov}--\eqref{eq:Sigma-cond}, in which $\alpha_s$ cancels:
\begin{equation}
\Gamma_{n,\perp}^{\rm higher}=\frac{\tau_2^2-\tau_1^2}{\ln(\tau_2/\tau_1)}\,\frac{K^{\rm wa}_{n,\perp}(\beta,\theta)}{\overline W_0(\beta,\theta)}\,\big[1+{\cal O}(\tau_2)\big] .
\label{eq:Gamma-high}
\end{equation}
The two coefficients differ in their normalization, inclusive in Eq.~\eqref{eq:Gamma-low} and conditional in Eq.~\eqref{eq:Gamma-high}. Their ratio, $\Gamma^{\rm higher}/\Gamma^{\rm low}=\pi(\tau_2^2-\tau_1^2)/[\alpha_s\overline W_0\,\tau_1^2\ln(\tau_2/\tau_1)]$, exceeds unity whenever the probability of a resolved emission in the shell, $(\alpha_s/\pi)\overline W_0\ln(\tau_2/\tau_1)$, is smaller than $(\tau_2^2-\tau_1^2)/\tau_1^2$; for the benchmarks below the ratio is $10$--$15$, so that the higher bin dephases more strongly. The ${\cal O}(\tau_2)$ correction in Eq.~\eqref{eq:Gamma-high} has two sources. The first is the next-to-soft correction to the emission rate in the normalization, from $m_{-1}m_0^\dagger$, which we evaluate. For $\tau_2=0.3$ it reduces both $\Gamma_n^{\rm higher}$ and $\Gamma_\perp^{\rm higher}$ by the same factor, by $16$--$28\%$ for the benchmarks below, and leaves their ratio unchanged. The second is the interference of $m_0$ with $m_1$ in the numerator, which is of the same relative order but not universal. We therefore quote the leading-power values of Eq.~\eqref{eq:Gamma-high} with an uncertainty of about $30\%$ from these next-to-soft corrections. A separate and larger effect, the emission of more than one wide-angle gluon, is estimated in Sec.~\ref{sec:multi-bins}. The square of the conditional mean displacement, which enters $|{\cal Q}|$ at this order (Sec.~\ref{sec:dephasing}), vanishes at $\theta=\pi/2$ by the symmetry $\theta\to\pi-\theta$ and is below $1\%$ of $\Gamma_n^{\rm higher}$ for the other benchmarks. Figure~\ref{fig:resolution} shows the resolution dependence of the cumulative coefficients and of $C_{nn}$ and $|{\cal Q}|$ in the higher bin. The $\tau^2$ scaling replaces the logarithmic dependence that one would obtain by combining an energy-independent recoil angle with the eikonal measure.

The resolution window is motivated as follows. The expansion parameters are $\tau=E_{\rm cut}/E$ and $E_{\rm cut}/p_T$, Eq.~\eqref{eq:window}, and not $E_{\rm cut}/Q$; the photon virtuality does not enter the kernels. We restrict to $\tau_2\le0.3$ and $E_{\rm cut}^{(2)}\le p_T/2$, which is the practical form of the condition $E_{\rm cut}\ll p_T$ in Eq.~\eqref{eq:window}, and require $E_{\rm cut}^{(1)}\gtrsim0.4$~GeV. As discussed above, the low edge is nevertheless exposed to non-perturbative corrections of comparable size. As benchmarks we take $(\tau_1,\tau_2)=(0.1,0.3)$ for $c\bar c$ pairs with $M=8$~GeV and $b\bar b$ pairs with $M=12$ and $16$~GeV, which corresponds to $E_{\rm cut}^{(1)}=0.4$--$0.8$~GeV and $E_{\rm cut}^{(2)}=1.2$--$2.4$~GeV, below $p_T\simeq3$--$6$~GeV. The ratio $E_{\rm cut}^{(2)}/p_T$ is $0.32$--$0.50$ (Table~\ref{tab:bench}); the two points with $E_{\rm cut}^{(2)}/p_T\ge0.45$, marked in Table~\ref{tab:bench}, lie at the edge of the window \eqref{eq:window} and carry a correspondingly larger uncertainty. In particular, the soft approximation is never extended to energies of order $Q$ or $E$.

\subsection{Spin-dephasing ratios}

In each bin we define the normal correlation $C_{nn}$, the rotation-invariant in-plane magnitude $|{\cal Q}|$ of Eq.~\eqref{eq:Qmag}, and the spin-dephasing ratio
\begin{equation}
    R_{\rm dep} = \frac{|{\cal Q}|}{|C_{nn}|}=\frac{\frac12\sqrt{(C_{rr}-C_{kk})^2+(C_{rk}+C_{kr})^2}}{|C_{nn}|} .
    \label{eq:Rdep_def}
\end{equation}
For the symmetric-traceless Born block this coincides, up to a constant normalization, with the ratio $\sqrt{C_{rr}^2+C_{kk}^2+2C_{rk}^2}/(2|C_{nn}|)$. For a general block, however, only Eq.~\eqref{eq:Rdep_def} is invariant under rotations about $\hat n$ and insensitive to the trace and antisymmetric parts. In the longitudinal channel $|{\cal Q}|=C_{nn}=R_{\rm dep}=1$ at Born level for every $(\beta,\theta)$. We form the radiation-binned single ratios and the double ratio
\begin{equation}
    {\mathcal D}_{nn}=\frac{C_{nn}^{\rm higher}}{C_{nn}^{\rm low}},
    \qquad
    {\mathcal D}_{\cal Q}=\frac{|{\cal Q}|^{\rm higher}}{|{\cal Q}|^{\rm low}},
    \qquad
    {\mathcal D}_{\rm EIC}^{\rm rad} = \frac{ R_{\rm dep}^{\rm higher}    }{   R_{\rm dep}^{\rm low}    } =\frac{{\cal D}_{\cal Q}}{{\cal D}_{nn}}\, ,
    \label{eq:D_EIC_rad}
\end{equation}
all evaluated at fixed Born kinematics (Sec.~\ref{sec:selection}). Using Eq.~\eqref{eq:Csoft},
\begin{equation}
{\cal D}_{nn}\simeq1-\Delta\Gamma_\perp,
\qquad
{\cal D}_{\cal Q}\simeq1-\Delta\Gamma_n-\frac{\Delta\Gamma_\perp}{2},
\qquad
    {\mathcal D}_{\rm EIC}^{\rm rad}   \simeq  \exp\left[\frac{\Delta\Gamma_\perp}{2}-\Delta\Gamma_n\right],
    \qquad \Delta\Gamma\equiv\Gamma^{\rm higher}-\Gamma^{\rm low}.
    \label{eq:D_EIC_prediction}
\end{equation}
Since $\Gamma^{\rm higher}>\Gamma^{\rm low}$, both ${\cal D}_{nn}$ and ${\cal D}_{\cal Q}$ are below unity. Since $K^{\rm wa}_\perp>2K^{\rm wa}_n$ (Sec.~\ref{sec:kernels}), the double ratio is above unity, ${\cal D}_{\rm EIC}^{\rm rad}>1$. A pure rotation channel would give ${\cal D}_{\rm EIC}^{\rm rad}=e^{-\Delta\Gamma_n}<1$, so the sign of ${\cal D}_{\rm EIC}^{\rm rad}-1$ discriminates between the two channels. The pair $({\cal D}_{nn},{\cal D}_{\cal Q})$ determines $\Delta\Gamma_\perp$ and $\Delta\Gamma_n$ separately. Indicative values for the benchmarks are listed in Table~\ref{tab:bench}. At leading power and with a single emission, the normal correlation in the higher bin is reduced by $2$--$4\%$, the in-plane magnitude by $1$--$2.5\%$, and ${\cal D}_{\rm EIC}^{\rm rad}-1\simeq0.5$--$0.9\%$; multiple emissions lower these deviations by about a third, Sec.~\ref{sec:multi-bins}. Thus the experimentally relevant signature is not merely a change of the heavy-flavour rate, but a radiation-dependent and anisotropic degradation of the spin-correlation tensor whose dominant component is the loss of normal correlation.

\begin{table*}[t]
\centering
\caption{Indicative values for the radiation bins of Eq.~\eqref{eq:bins} with $(\tau_1,\tau_2)=(0.1,0.3)$, $\alpha_s=0.25$ and $R=R_B=0.5$. $E^{(1)}_{\rm cut}$--$E^{(2)}_{\rm cut}$ and $p_T=E\beta\sin\theta$ are in GeV. The higher-bin coefficients are the leading-power values of Eq.~\eqref{eq:Gamma-high}, with an ${\cal O}(\tau_2)$ uncertainty of about $30\%$; the calculable part of this correction, from the normalization, lowers both by $16$--$28\%$ (Sec.~\ref{sec:resolution}). $N_{3\sigma}$ is the number of longitudinally separated, fixed-Born double-analyzer events, in two equally populated bins, needed for a $3\sigma$ measurement of $1-{\cal D}_{nn}$ with the moment estimator of Eq.~\eqref{eq:Nreq} and the ideal decay-only dilution $\kappa=0.57$; for $\kappa=0.2$ it is larger by a factor of $8.4$. The number of recorded events is larger by the separation factor of Eq.~\eqref{eq:LT-penalty}. For the $b\bar b$ rows, $\kappa$ is that of the $\Lambda_c$ analyzer and serves only for comparison. Benchmarks marked $^\dagger$ lie at the edge of the window \eqref{eq:window}. The effect of multiple wide-angle emissions is shown in Table~\ref{tab:multi}.}
\label{tab:bench}
\begin{tabular}{ccccccccccccccccc}
\hline\hline
pair & $M$ & $\beta$ & $\theta$ & $E^{(1)}_{\rm cut}$--$E^{(2)}_{\rm cut}$ & $p_T$ & $E^{(2)}_{\rm cut}/p_T$ & $K^{\rm wa}_n$ & $K^{\rm wa}_\perp$ & $\Gamma_n^{\rm low}$ & $\Gamma_\perp^{\rm low}$ & $\Gamma_n^{\rm higher}$ & $\Gamma_\perp^{\rm higher}$ & ${\cal D}_{nn}$ & ${\cal D}_{\cal Q}$ & ${\cal D}^{\rm rad}_{\rm EIC}$ & $N_{3\sigma}$\\
\hline
$c\bar c$ & 8 & 0.93 & $\pi/2$ & 0.4--1.2 & 3.7 & 0.32 & 0.32 & 2.27 & $2.6\!\times\!10^{-4}$ & $1.8\!\times\!10^{-3}$ & 0.0027 & 0.0186 & 0.983 & 0.989 & 1.0061 & $3.4\!\times\!10^{6}$ \\
$c\bar c$ & 8 & 0.93 & 1.0 & 0.4--1.2 & 3.1 & 0.38 & 0.47 & 2.72 & $3.8\!\times\!10^{-4}$ & $2.2\!\times\!10^{-3}$ & 0.0046 & 0.0261 & 0.976 & 0.984 & 1.0080 & $1.7\!\times\!10^{6}$ \\
$b\bar b$ & 16 & 0.80 & $\pi/2$ & 0.8--2.4 & 6.4 & 0.37 & 0.58 & 2.29 & $4.6\!\times\!10^{-4}$ & $1.8\!\times\!10^{-3}$ & 0.0058 & 0.0231 & 0.979 & 0.984 & 1.0054 & $2.1\!\times\!10^{6}$ \\
$b\bar b$ & 16 & 0.80 & 1.0 & 0.8--2.4 & 5.4 & 0.45$^\dagger$ & 0.77 & 2.85 & $6.1\!\times\!10^{-4}$ & $2.3\!\times\!10^{-3}$ & 0.0089 & 0.0328 & 0.969 & 0.976 & 1.0073 & $1.0\!\times\!10^{6}$ \\
$b\bar b$ & 12 & 0.60 & $\pi/2$ & 0.6--1.8 & 3.6 & 0.50$^\dagger$ & 0.76 & 3.07 & $6.0\!\times\!10^{-4}$ & $2.4\!\times\!10^{-3}$ & 0.0091 & 0.0369 & 0.965 & 0.974 & 1.0091 & $8.1\!\times\!10^{5}$ \\
\hline\hline
\end{tabular}
\end{table*}

\subsection{Multiple wide-angle emissions in the radiation bins}
\label{sec:multi-bins}

The single-emission treatment of the higher bin requires the probability of a wide-angle emission in the shell, $(\alpha_s/\pi)\overline W_0\ln(\tau_2/\tau_1)$, to be small. For the benchmarks this probability is $0.5$--$0.8$, since $\overline W_0\simeq6$--$9$ for $R_B=0.5$, with the initial-final dipoles of Eq.~\eqref{eq:W0} carrying the largest color factors. Configurations with two or more wide-angle emissions that add up to an $E_{\rm rad}$ inside the shell are therefore not suppressed. We estimate their effect within the independent-emission model of Sec.~\ref{sec:multiple}: wide-angle emissions form a Poisson process in energy with intensity
\begin{equation}
d\nu=a\,\frac{d\omega}{\omega},\qquad a\equiv\frac{\alpha_s}{\pi}\,\overline W_0 ,
\label{eq:poisson}
\end{equation}
and the decohering Gram matrices of the individual emissions add. By Eq.~\eqref{eq:Sigma-cond}, an emission of energy $\omega$ contributes $2(\omega/E)^2K^{\rm wa}/\overline W_0$ to $\Gamma_{n,\perp}$, so that the coefficient of a bin is determined by the conditional expectation of $\sum_i\omega_i^2$ at fixed $E_{\rm rad}=\sum_i\omega_i$.

For a scale-invariant Poisson process the density of the total energy below the upper end of the soft spectrum is $p(E)=C\,E^{a-1}$, with a constant $C$ that drops out of all conditional averages. The Mecke identity for Poisson processes gives
\begin{equation}
\Big\langle\sum_i\omega_i^2\,\delta\big(E_{\rm rad}-S\big)\Big\rangle
=\int_0^S a\,\frac{d\omega}{\omega}\,\omega^2\,p(S-\omega)
=a\,C\,S^{a+1}\,{\rm B}(2,a)
=\frac{S^2}{1+a}\,p(S),
\label{eq:mecke}
\end{equation}
with ${\rm B}$ the Euler beta function, so that
\begin{equation}
\Big\langle\sum_i\omega_i^2\Big\rangle_{E_{\rm rad}=S}=\frac{S^2}{1+a}
\label{eq:cond-sum}
\end{equation}
exactly within the model. A single emission corresponds to $a\to0$. Averaging Eq.~\eqref{eq:cond-sum} over a bin with the density $p(S)\propto S^{a-1}$ gives
\begin{equation}
\Gamma^{\rm low}_{n,\perp}(a)=\frac{2}{(1+a)(2+a)}\,\Gamma^{\rm low}_{n,\perp},
\qquad
\Gamma^{\rm higher}_{n,\perp}(a)=F(a)\,\Gamma^{\rm higher}_{n,\perp},
\qquad
F(a)=\frac{2a\,\ln(\tau_2/\tau_1)}{(1+a)(2+a)}\,
\frac{\tau_2^{2+a}-\tau_1^{2+a}}{(\tau_2^a-\tau_1^a)(\tau_2^2-\tau_1^2)} ,
\label{eq:multi-factors}
\end{equation}
where $\Gamma^{\rm low}$ and $\Gamma^{\rm higher}$ without argument are the single-emission values \eqref{eq:Gamma-low} and \eqref{eq:Gamma-high}. Both factors tend to unity as $a\to0$, and $F$ depends on $\tau_1$ and $\tau_2$ only through their ratio.

Three features of Eq.~\eqref{eq:multi-factors} matter for the observable. First, the same factor multiplies $\Gamma_n$ and $\Gamma_\perp$ within each bin, so the ratio $\Gamma_\perp/\Gamma_n=K^{\rm wa}_\perp/K^{\rm wa}_n$, and with it the sign of ${\cal D}^{\rm rad}_{\rm EIC}-1$, is unchanged. Second, both coefficients decrease: additional emissions share the measured energy, and the decoherence grows with the square of each individual energy. Third, the correction is large. With the values of $\overline W_0$ that enter Table~\ref{tab:bench} through Eq.~\eqref{eq:Gamma-high}, one has $a=0.48$--$0.71$ and $F=0.66$--$0.74$, and the deviations from unity of ${\cal D}_{nn}$, ${\cal D}_{\cal Q}$ and ${\cal D}^{\rm rad}_{\rm EIC}$ are reduced by about a third (Table~\ref{tab:multi}). The model neglects correlated emission, color evolution and non-global configurations, so the difference between Tables~\ref{tab:bench} and \ref{tab:multi} is an estimate of the size of the multiple-emission correction rather than a prediction of it.

\begin{table}[t]
\centering
\caption{Independent-emission estimate of multiple wide-angle emissions, Eqs.~\eqref{eq:poisson}--\eqref{eq:multi-factors}, for the benchmarks of Table~\ref{tab:bench}. $\overline W_0$ is the value entering Eq.~\eqref{eq:Gamma-high}, $a=(\alpha_s/\pi)\overline W_0$ with $\alpha_s=0.25$, and $2/[(1+a)(2+a)]$ and $F$ rescale the low-bin and higher-bin coefficients. The last three columns replace the corresponding single-emission values of Table~\ref{tab:bench}.}
\label{tab:multi}
\begin{tabular}{cccccccccc}
\hline\hline
pair & $M$ & $\theta$ & $\overline W_0$ & $a$ & $\frac{2}{(1+a)(2+a)}$ & $F$ & ${\cal D}_{nn}$ & ${\cal D}_{\cal Q}$ & ${\cal D}^{\rm rad}_{\rm EIC}$\\
\hline
$c\bar c$ & 8 & $\pi/2$ & 8.9 & 0.71 & 0.43 & 0.66 & 0.988 & 0.993 & 1.0042\\
$c\bar c$ & 8 & 1.0 & 7.6 & 0.60 & 0.48 & 0.70 & 0.983 & 0.988 & 1.0056\\
$b\bar b$ & 16 & $\pi/2$ & 7.2 & 0.57 & 0.49 & 0.70 & 0.985 & 0.989 & 1.0039\\
$b\bar b$ & 16 & 1.0 & 6.3 & 0.50 & 0.53 & 0.73 & 0.977 & 0.983 & 1.0053\\
$b\bar b$ & 12 & $\pi/2$ & 6.1 & 0.48 & 0.54 & 0.74 & 0.974 & 0.981 & 1.0067\\
\hline\hline
\end{tabular}
\end{table}

\subsection{Relation to the collinear channel}
\label{sec:collinear}

Unresolved quasi-collinear radiation from the heavy quarks also modifies the spin state, through the massive splitting matrix \cite{Catani:2000ef,Catani:2002hc}. Aoude \textit{et al.} treat it as a Kraus map \cite{Aoude:2025ovu,Aoude:2026eeg}. The heavy-quark mass regulates the collinear singularity at $k_\perp\sim(1-z)m$, but it does not attach a factor $\tau^2$ to this map. Parametrically,
\begin{equation}
\delta S_L^{\rm coll}\sim\frac{\alpha_s C_F}{2\pi}\,K^{\rm coll}\!\left(\frac{m}{E},R,z_{\rm cut}\right),
\qquad
\delta S_L^{\rm wa}=\frac{\alpha_s}{\pi}\,\tau_c^2\,(K^{\rm wa}_n+K^{\rm wa}_\perp),
\label{eq:coll-compare}
\end{equation}
with $K^{\rm coll}={\cal O}(1)$. For $\tau_c=0.2$ and $K^{\rm wa}_n+K^{\rm wa}_\perp\simeq3$--$4$ in the central region, the wide-angle term is $\simeq0.05\,\alpha_s$, whereas the collinear term is of order $C_F\alpha_s/(2\pi)\simeq0.2\,\alpha_s$. The wide-angle next-to-soft contribution is therefore not dominant in an inclusive measurement of the spin state, and the dead cone ($\theta_m=m/E\simeq0.4$ for the $c\bar c$ benchmarks at $E=4$~GeV, and $0.6$--$0.8$ for the $b\bar b$ benchmarks at $E=6$--$8$~GeV) does not by itself establish soft-recoil dominance.

The radiation-binned construction is designed so that this does not matter at the order considered. Emission inside the cones or along the beam does not contribute to $E_{\rm rad}$. At leading power the collinear and hard-virtual corrections factorize from the wide-angle soft emission, so that the state in either bin is the composition of a common resolution-independent map ${\cal E}_{\rm hc}$ with the wide-angle soft map of that bin. This is a leading-power factorization assumption, not a theorem for spin-dependent observables, and we make its consequences explicit. For the normal correlation, the mirror-odd map \eqref{eq:rho-odd}, applied to an arbitrary input state $\rho$, gives $C_{nn}\to(1-\Gamma_\perp)\,C_{nn}-2\,{\rm Im}\,\gamma^\perp_{kr}\,\bar P_n$, as follows from $\sigma_a\sigma_n\sigma_a=-\sigma_n$ and Eq.~\eqref{eq:pauli-triad}, where $\bar P_n$ is the antiquark polarization along $\hat n$. The mirror-even map leaves $C_{nn}$ unchanged. Writing the state in a bin as ${\cal E}_{\rm bin}\circ{\cal E}_{\rm hc}(\rho^{(0)})$, with ${\cal E}_{\rm hc}$ the common hard-collinear map, one therefore has $\ln C^{\rm bin}_{nn}=\ln C^{\rm hc}_{nn}-\Gamma^{\rm bin}_\perp+{\cal O}(\Gamma^2,\,\Gamma\,\bar P_n)$, and the hard-collinear factor cancels in ${\cal D}_{nn}$. For $|{\cal Q}|$ the same holds to first order in the deviation $\delta_{\rm hc}$ of ${\cal E}_{\rm hc}(\rho^{(0)})$ from the Born state. To first order the logarithms of the ratios in Eq.~\eqref{eq:D_EIC_rad} are thus additive in the two maps, and ${\cal E}_{\rm hc}$ cancels:
\begin{equation}
\ln{\cal D}=\ln{\cal D}\big|_{\rm wa}+{\cal O}\big(\alpha_s\,\Gamma\big).
\label{eq:coll-cancel}
\end{equation}
The wide-angle recoil is thus isolated by the measurement projection, not by dead-cone dominance. Radiation just outside the cones is part of $\Omega_{\rm wa}$ and is included in $K^{\rm wa}_{n,\perp}$, which retain the full dependence on $R$ and $m/E$. Configurations with one collinear and one wide-angle emission enter at ${\cal O}(\alpha_s^2)$.

The two bins start at different orders in $\alpha_s$: the low bin at the Born level, the higher bin with one real emission. For the normalized spin observables the counting is as follows. In the low bin the deviation from the Born state is ${\cal O}(\alpha_s\tau_1^2)$ from wide-angle emission and ${\cal O}(\alpha_s)$ from the hard virtual and collinear corrections. In the higher bin it is ${\cal O}(\tau_2^2)$ from the resolved emission, in which $\alpha_s$ cancels, and again ${\cal O}(\alpha_s)$ from the hard and collinear corrections. At leading soft power the one-loop soft current is spin independent, so the hard virtual correction acts on the Born spin state in the same way in both bins and cancels in the ratios. What remains are three residual terms. (a) The non-commutativity of the hard-collinear and soft maps, of relative order $\delta_{\rm hc}$; with $\delta_{\rm hc}\sim\alpha_s C_FK^{\rm coll}/(2\pi)$ from Eq.~\eqref{eq:coll-compare} this is a few per cent of $\Delta\Gamma$, comparable to the estimate $\alpha_s/\pi\simeq8\%$. (b) Terms proportional to the single-spin polarization $\bar P_n$ of the hard-collinear state, which vanishes at tree level and is generated only by absorptive parts at one loop, so that these terms are ${\cal O}(\alpha_s\Gamma)$ and further suppressed. (c) Configurations with one collinear and one wide-angle emission, which are of the same order as (a). This is the accuracy of Eq.~\eqref{eq:coll-cancel} for the hard and collinear corrections. It does not include the multiple wide-angle emissions of Sec.~\ref{sec:multi-bins}, which are a larger effect.

\subsection{Longitudinal-transverse separation}
\label{sec:LT}

The baseline $C_{nn}=1$ and the map \eqref{eq:Csoft} refer to the longitudinal photon. After division by the transverse flux and a full azimuthal integration about the photon axis, which removes the $LT$ and $TT$ interference harmonics, the measured spin-weighted and unpolarized cross sections are
\begin{equation}
N_{ij}(\epsilon)=\sigma_T\,C^T_{ij}+\epsilon\,\sigma_L\,C^L_{ij},
\qquad
D(\epsilon)=\sigma_T+\epsilon\,\sigma_L,
\qquad
C^{\rm meas}_{ij}=f_LC^L_{ij}+(1-f_L)\,C^T_{ij},
\qquad f_L=\frac{\epsilon\sigma_L}{\sigma_T+\epsilon\sigma_L},
\label{eq:LT}
\end{equation}
where $\epsilon$ is the photon polarization parameter. The transverse contribution is large. For the Born process at $M=10$~GeV we find $\sigma_L/\sigma_T\simeq0.02$--$0.06$ at $Q^2=4$~GeV$^2$, $0.07$--$0.2$ at $Q^2=16$~GeV$^2$, and $0.14$--$0.5$ at $Q^2=50$~GeV$^2$ over $0.6\le\beta\le0.8$, with $C^T_{nn}\simeq-(0.3$--$0.6)$ and $|{\cal Q}^T|\simeq0.14$--$0.3$. An unseparated sample therefore has $C_{nn}^{\rm meas}-1=(1-f_L)(C^T_{nn}-1)$, far from the longitudinal baseline. Between two radiation bins it acquires
\begin{equation}
\Delta C^{\rm meas}_{ij}=f_L\,\Delta C^L_{ij}+(1-f_L)\,\Delta C^T_{ij}+\Delta f_L\,\big(C^L_{ij}-C^T_{ij}\big),
\label{eq:LT-diff}
\end{equation}
where the last term arises if the radiation selection changes the $L/T$ composition. An unseparated double ratio is thus dominated by the transverse channel, whose radiative decoherence is not computed here.

The longitudinal tensor is obtained without a model of the transverse density matrix by a spin-dependent Rosenbluth separation. One takes data at two or more values of $\epsilon$, obtained from different beam-energy combinations at fixed $(x,Q^2,\Phi_B)$ and in each radiation bin, and forms
\begin{equation}
C^L_{ij}=\frac{N_{ij}(\epsilon_2)-N_{ij}(\epsilon_1)}{D(\epsilon_2)-D(\epsilon_1)} .
\label{eq:rosenbluth}
\end{equation}
The separation must be performed in each radiation bin before the ratios \eqref{eq:D_EIC_rad} are formed, which also removes the $\Delta f_L$ term. With more than two settings, $N_{ij}$ and $D$ are fitted linearly in $\epsilon$, which tests the one-photon and acceptance assumptions. The statistical error grows as $|\epsilon_2-\epsilon_1|^{-1}$. Writing $N_{ij}=D\,C^{\rm meas}_{ij}$, using $D(\epsilon_2)-D(\epsilon_1)=\Delta\epsilon\,\sigma_L$, and splitting a total sample equally between the two settings, error propagation gives $\delta C^L_{ij}\simeq2\,[\bar\epsilon/(\bar f_L\Delta\epsilon)]\,\delta C_{ij}$, where $\delta C_{ij}$ is the error of the same total sample if it were purely longitudinal and $\bar f_L$ is the longitudinal fraction at $\bar\epsilon=(\epsilon_1+\epsilon_2)/2$. The number of recorded events therefore exceeds that of a purely longitudinal sample by
\begin{equation}
\frac{N_{\rm rec}}{N_L}\simeq4\left(\frac{\bar\epsilon}{\bar f_L\,\Delta\epsilon}\right)^2 ,
\label{eq:LT-penalty}
\end{equation}
which for $\bar f_L=0.1$--$0.3$ and $\bar\epsilon/\Delta\epsilon=2$--$3$ lies between about $2\times10^2$ and $4\times10^3$. If an imperfect separation leaves a transverse fraction $r_T\ll1$, then $C^{\rm ext}_{nn}-1=r_T(C^T_{nn}-1)+\delta C^{L,{\rm rad}}_{nn}$. With $C_{nn}^T-1\simeq-1.4$, the residual must satisfy $|\Delta r_T|\lesssim3\times10^{-3}$ between the bins for the contamination to remain below a quarter of the predicted $\Delta\Gamma_\perp$.

\subsection{Born-kinematic selection and production-plane resolution}
\label{sec:selection}

$E_{\rm rad}$ is correlated with the hard kinematics, so the two bins populate different distributions $f_{\rm bin}(\Phi_B)=f_0+\delta f_{\rm bin}$ of $\Phi_B=(x,Q^2,M,\beta,\theta)$. For a bin-averaged observable,
\begin{equation}
\Delta\langle O\rangle=\int d\Phi_B\,f_0\,\Delta\delta O
+\int d\Phi_B\,\big[\delta f_{\rm higher}-\delta f_{\rm low}\big]\,O^{(0)}(\Phi_B)+{\cal O}(\alpha_s^2) .
\label{eq:selection}
\end{equation}
In the longitudinal channel $O^{(0)}=C_{nn}=|{\cal Q}|=R_{\rm dep}=1$ at every Born point, so the Born-level selection term in Eq.~\eqref{eq:selection} vanishes identically for the observables \eqref{eq:D_EIC_rad}. A residual of order $\int[\delta f_{\rm higher}-\delta f_{\rm low}]\,\Gamma(\Phi_B)$ remains, because the coefficients depend strongly on $\beta\sin\theta$. It is removed by forming the ratios in fixed Born bins and then averaging with a common weight $w(\Phi_B)$ chosen independently of the radiation bin, which is equivalent to unfolding or reweighting both samples to the same Born distribution. In finite Born bins a residual arises from a mismatch of the bin centroids or widths, $\delta O\simeq\Delta\bar\Phi^a\partial_aO+\frac12\Delta V^{ab}\partial_a\partial_bO$, which is again of order $\Gamma$ times the mismatch. A second residual arises because $|{\cal Q}|$ is evaluated on the bin-averaged tensor, and $|\langle{\cal Q}\rangle|\neq\langle|{\cal Q}|\rangle$. A coherent mean rotation $\langle\delta\phi_n\rangle(\Phi_B)$ that varies across a Born bin dephases the averaged tensor by $2\,{\rm Var}_{\rm bin}\langle\delta\phi_n\rangle$. In the higher bin the mean rotation is ${\cal O}(\tau_2)$ rather than ${\cal O}(\alpha_s\tau_1)$, so this term does not cancel in the ratios; for a $\theta$ bin of width $\Delta\theta$ it is $(\partial_\theta\langle\delta\phi_n\rangle)^2(\Delta\theta)^2/6$, and the Born-bin widths must be chosen small enough for it to remain below $\Delta\Gamma_n$. At $\theta=\pi/2$ the mean rotation vanishes by symmetry, but its derivative does not.

Finite resolution on the reconstructed production plane acts on the spin tensor like a rotation. A misreconstruction of the in-plane axes by $\delta\psi$ (a rotation about $\hat n$) multiplies ${\cal Q}$ by $R_2=\langle e^{2i\delta\psi}\rangle$, which is $e^{-2\sigma_\psi^2}$ for a Gaussian, and leaves $C_{nn}$ invariant. A tilt of the reconstructed plane by $\delta\psi_\perp$ about an in-plane axis reduces $C_{nn}$ and $|{\cal Q}|$ at ${\cal O}(\sigma_{\psi_\perp}^2)$. These effects mimic $\Gamma_n$ and $\Gamma_\perp$, respectively. They are separated as follows. (i) The plane response is a detector property that does not involve the spin, so it can be determined from the reconstructed-versus-true directions in simulation, validated with data, separately in each radiation bin. The measured tensors are corrected as ${\cal Q}^{\rm corr}={\cal Q}^{\rm obs}/R_2$, or more generally unfolded with a response matrix acting on all spin moments. (ii) Only the bin dependence matters: ${\cal D}^{\rm obs}={\cal D}^{\rm phys}\,R_2^{\rm higher}/R_2^{\rm low}$, so a common resolution cancels. The requirement is that the bin-to-bin change of the response be known to better than the signal, for instance $2|\sigma_\psi^2({\rm higher})-\sigma_\psi^2({\rm low})|\ll\Delta\Gamma_n\simeq0.005$, that is, $|\Delta\sigma_\psi^2|\ll2.5\times10^{-3}$~rad$^2$, and similarly $|\Delta\sigma_{\psi_\perp}^2|\ll\Delta\Gamma_\perp\simeq0.02$. (iii) The physical coefficients have a definite kinematic dependence, $\Gamma\propto\tau^2$ and $\Gamma_\perp\propto(\beta\sin\theta)^{-2}$, whereas the resolution effects follow the angular resolution of the reconstructed axes. A correlation of the reconstruction quality with the radiation activity is reduced by defining the axes from the heavy-flavor hadrons and by excluding the jet regions from $E_{\rm rad}$.

\subsection{Analyzing powers, spin transfer and statistical reach}
\label{sec:stat}

In the ratios \eqref{eq:D_EIC_rad} the weak-decay analyzing powers cancel exactly when the same decay channels, and their charge conjugates, are used in both bins. In Eq.~\eqref{eq:spin_analyzer_distribution}, however, the measured correlations are $\kappa_a\kappa_b\,A_{ij}\,C_{ij}$, where $A_{ij}$ contains acceptance, efficiency and migration. The spin transfer ${\cal T}$ and $A_{ij}$ need not be the same in the two bins. Writing $\kappa^{\rm higher}A^{\rm higher}/(\kappa^{\rm low}A^{\rm low})=1+\delta_\kappa$, one has ${\cal D}^{\rm obs}-1=({\cal D}-1)+\delta_\kappa$, so the cancellation requires $|\delta_\kappa|\ll|\Delta\Gamma|$. The spin transfer to $\Lambda_c$ ($\Lambda_b$) is reduced by feed-down from $\Sigma_{c}^{(*)}$ ($\Sigma_{b}^{(*)}$) \cite{Falk:1993rf}, and it may depend on the hadronic activity near the heavy quark. Its bin dependence is not known. It is a required calibration, to be determined for example from the single-baryon polarization in the two bins; excluding the heavy-flavor cones from $E_{\rm rad}$ is expected to reduce it. Where $\delta_\kappa$ cannot be controlled, the two samples must be unfolded with separate response matrices before the ratios are formed.

Decay analyzers do not measure the spin projections directly. From Eq.~\eqref{eq:spin_analyzer_distribution}, integrated over the analyzer directions, $\langle\hat a_i\hat b_j\rangle=\kappa C_{ij}/9$ and $\langle\hat a_i^2\hat b_j^2\rangle=1/9$, so that the moment estimator $\widehat C_{ij}=9\,\hat a_i\hat b_j/\kappa$ has the per-event variance $(9-\kappa^2C_{ij}^2)/\kappa^2$. For two bins with fractions $f_{\rm h}$ and $f_{\rm l}=1-f_{\rm h}$ of the events, a $Z\sigma$ measurement of a difference $\Delta C$ requires
\begin{equation}
N_{\rm tot}\gtrsim\frac{Z^2\,(9-\kappa^2)}{\kappa^2(\Delta C)^2}\left(\frac{1}{f_{\rm h}}+\frac{1}{f_{\rm l}}\right)
\label{eq:Nreq}
\end{equation}
longitudinally separated, fixed-Born double-analyzer events, where we have set $C\simeq1$; for equal populations the last factor is $4$. With $\Lambda_c^+\to\Lambda\pi^+$, $\alpha=-0.755\pm0.005\pm0.003$ \cite{LHCb:2024tnq}, the ideal decay-only dilution is $\kappa=\alpha^2=0.57$, and the effective value with spin transfer included is smaller; we take $\kappa=0.2$ as a representative value. The predicted single-emission $1-{\cal D}_{nn}\simeq0.017$--$0.035$ then requires $N_{\rm tot}\simeq(0.8$--$3.4)\times10^6$ events for $\kappa=0.57$ and $(0.7$--$2.9)\times10^7$ for $\kappa=0.2$, and the multiple-emission estimate of Table~\ref{tab:multi} increases these numbers by a factor of about two. The smaller deviation ${\cal D}_{\rm EIC}^{\rm rad}-1$ requires $\gtrsim10^7$ events for $\kappa=0.57$ and $\gtrsim10^8$ for $\kappa=0.2$. The $b\bar b$ entries of Table~\ref{tab:bench} use the $\Lambda_c$ dilution for comparison only; a $\Lambda_b$ analyzer has a different analyzing power and a much smaller yield. The number of recorded events exceeds $N_{\rm tot}$ by the separation factor \eqref{eq:LT-penalty}. The available yield before longitudinal-transverse separation is
\begin{equation}
N_{\rm acc}={\cal L}_{\rm int}\,\sigma_{Q\bar Q}\,\epsilon_{\rm sel}\,
\Big[f(c\to\Lambda_c^+)\,{\cal B}(\Lambda_c^+\to\Lambda\pi^+)\,{\cal B}(\Lambda\to p\pi^-)\,\epsilon_{\rm rec}\Big]^2 ,
\label{eq:Nacc}
\end{equation}
where the bracket is $0.06\times0.013\times0.64\simeq5\times10^{-4}$ per baryon \cite{Lisovyi:2015uqa,ParticleDataGroup:2024cfk} before reconstruction efficiency. With an integrated luminosity of $100$~fb$^{-1}$ \cite{AbdulKhalek:2021gbh} and, for illustration, $\sigma_{c\bar c}=10$~nb in the DIS region, the double-$\Lambda_c$ yield is at most ${\cal O}(10^2)\,\epsilon_{\rm rec}^2$ before the bin populations and the fixed-Born selection are applied. Together with the factor \eqref{eq:LT-penalty}, this is about six or more orders of magnitude below the requirement \eqref{eq:Nreq}. Higher-rate decays, such as $\Lambda_c^+\to pK^-\pi^+$ with a full amplitude-based polarimeter, improve this by at most one to two orders of magnitude. The effect is therefore not measurable with exclusive heavy-baryon pairs at nominal EIC luminosities. Its observation would require substantially larger samples or more efficient heavy-quark spin analyzers, and the radiation-binned ratios should be regarded as a theoretically specified target.

\section{Conclusion and outlook}
\label{sec:conclusion-clean}

In this work we have identified a mechanism by which QCD radiation can reduce the spin entanglement of a heavy $q\bar q$ pair. At leading order, the longitudinal photon channel produces a pure two-spin state with a fixed correlation plane. Since there is no final-state radiation at this order, the spin density matrix remains pure, and the longitudinal channel displays maximal entanglement. The normal correlation $C_{nn}=1$ is enforced by the mirror symmetry of the production plane. \\

At next-to-leading order the situation changes qualitatively. Real soft-gluon emission enlarges the final-state Hilbert space. When the gluon is unresolved, the observed system is no longer the full quantum state, but the reduced density matrix obtained after summing over the momentum, color and polarization of the unobserved gluon. The leading eikonal soft factor of this radiation is not itself responsible for spin decoherence: it is scalar in spin space and cancels from normalized spin observables, together with the spin-independent virtual correction. The first genuine decoherence is tied to the next-to-soft amplitude. Relative to the emitted state it displaces the spin state by an amount of order $\omega/E$, so the decohering contribution of an unresolved region scales as $\alpha_s(E_{\rm cut}/E)^2$, with a positive Gram kernel that includes the exact color-octet structure. \\

The central result is that the decoherence has two components, separated by the mirror parity $\Pi_n=\sigma_n\otimes\sigma_n$ of the production plane. The mirror-odd part of the radiation, generated by the out-of-plane recoil, by the incoming-gluon polarization normal to the plane and by intrinsic-spin insertions, flips the normal-spin parity of the pair and reduces $C_{nn}$ by $\Gamma_\perp$. It is the dominant channel: its kernel exceeds that of the rotation channel by a factor of at least $3.5$, grows as the inverse square of the heavy-quark transverse momentum, and reflects the event-by-event tilt of the hard-scattering plane relative to the reconstructed production plane, which cannot be followed by any axes built from the observed momenta. The mirror-even part of the radiation produces small event-by-event rotations of the spin-correlation plane about its normal. After the trace these become a dephasing channel with Lindblad operator $J_n=\sigma_n\otimes I+I\otimes\sigma_n$, Eq.~\eqref{eq:rhoav}, which suppresses the charge-two in-plane coherence by $1-\Gamma_n$ at fixed order (by $e^{-\Gamma_n}$ for independent emissions) and leaves the normal correlation unchanged. Both coefficients follow in closed form from the complete next-to-soft amplitude, Eqs.~\eqref{eq:Kclosed}--\eqref{eq:Kpolys}, with the analytic limits \eqref{eq:Kthreshold} and \eqref{eq:Kultrarel}. The normal correlation is therefore protected only within the rotation channel. \\

In summary, the longitudinal photon channel provides an unusually clean arena in which QCD spin entanglement is produced maximally at leading order and degraded in a controlled and calculable way by unresolved radiation. Subleading soft recoil produces normal-spin flips and stochastic rotations of the heavy-quark spin-correlation structure, and tracing over the unobserved gluon converts them into decoherence. The resulting anisotropy, a stronger loss of normal correlation than of in-plane coherence, is measured by the radiation-binned ratios ${\cal D}_{nn}$, ${\cal D}_{\cal Q}$ and ${\cal D}^{\rm rad}_{\rm EIC}$, formed at fixed Born kinematics after longitudinal-transverse separation. Provided the analyzing powers, acceptances and the production-plane response are controlled bin by bin, collinear and hard-virtual effects and the Born-kinematic selection cancel in these ratios. For the benchmarks the normal correlation in the higher-radiation bin is reduced by $2$--$4\%$ and ${\cal D}^{\rm rad}_{\rm EIC}-1\simeq0.5$--$0.9\%$ at leading power with a single emission, with an uncertainty of about $30\%$ from next-to-soft corrections to the conditional state. Multiple wide-angle emissions, estimated in closed form in an independent-emission model, lower these deviations to $1$--$2.6\%$ and $0.4$--$0.7\%$ without changing the sign of ${\cal D}^{\rm rad}_{\rm EIC}-1$; the non-perturbative content of the radiation variable at its low edge is a further correction that must be calibrated. Measuring these effects with exclusive heavy-baryon pairs would require samples about six or more orders of magnitude beyond the yield at nominal EIC luminosity. Natural extensions are the transverse-photon channel, the resummation of non-global logarithms of the radiation resolution, and the sub-subleading soft amplitude, which would complete the prediction for the individual spin-correlation components.
 \\

\section{Appendix A: Angular and energy integrations, and Pauli algebra}
\label{app:angular-clean}

\paragraph*{Energy integrals.} With $g_s^2\,d^3k/[(2\pi)^32\omega]=(\alpha_s/\pi)\,\omega\,d\omega\,d\Omega/(4\pi)$, the three terms of the real-emission matrix integrate as
\begin{equation}
\int\omega\,d\omega\,\frac{1}{\omega^2}=\ln\frac{\omega_b}{\omega_a},
\qquad
\int\omega\,d\omega\,\frac{1}{\omega}=\omega_b-\omega_a,
\qquad
\int\omega\,d\omega=\frac{\omega_b^2-\omega_a^2}{2},
\label{eq:energy-ints}
\end{equation}
for the eikonal, interference and next-to-soft terms, respectively. For the cumulative region $0\le\omega<E_{\rm cut}$ the last term gives $E_{\rm cut}^2/2$ and hence Eq.~\eqref{eq:Kdef}. For the shell, the ratio of the last and first terms gives Eq.~\eqref{eq:Sigma-cond}.

\paragraph*{Octet eikonal kernel in the pair rest frame.} With the photon along $+\hat z$, the incoming gluon along $-\hat z$, $\hat\ell$ the soft-gluon direction, $\hat c=\hat k\cdot\hat\ell$, $d_\pm=1\mp\beta\hat c$ and $d_g=1+\hat\ell\cdot\hat z$, Eq.~\eqref{eq:W0} becomes
\begin{equation}
\omega^2W_0=C_A\left[\frac{1+\beta\cos\theta}{d_+d_g}+\frac{1-\beta\cos\theta}{d_-d_g}\right]
-\frac{1}{N_c}\,\frac{1+\beta^2}{d_+d_-}
-C_F\left(1-\beta^2\right)\left(\frac{1}{d_+^2}+\frac{1}{d_-^2}\right),
\label{eq:W0-closed}
\end{equation}
and $\overline W_0=\int_{\Omega_{\rm wa}}\frac{d\Omega}{4\pi}\,\omega^2W_0$. The beam cone is required for $\overline W_0$, since $\omega^2W_0$ has a non-integrable collinear singularity at $d_g=0$.

\paragraph*{Closed-form next-to-soft kernels.} The derivation of Eqs.~\eqref{eq:Kclosed}--\eqref{eq:Kpolys} proceeds in four exact steps. (i) In units $E=1$, with $h\equiv Q^2/M^2$ kept as a symbol and the rational parametrizations $\beta=2u/(1+u^2)$, $\sqrt{1-\beta^2}=(1-u^2)/(1+u^2)$ and $v=\tan(\theta/2)$, every ingredient of the eight diagrams of Appendix~B is a rational function of $(u,v,h,\hat\ell,\omega)$. The photon and target momenta follow from $q+p=p_1+p_2+k$, $q^2=-Q^2$, $p^2=0$ with the photon along $+\hat z$. This gives $q=(1-h,0,0,1+h)+\frac{\omega}{2}(1+\hat\ell_z)(1+h,0,0,1-h)+O(\omega^2)$, with the $O(\omega^2)$ terms fixed by the same conditions, and $\epsilon_L\propto(q^3,0,0,q^0)$. Expanding each diagram in $\omega$, the poles $p_{q,\bar q}\cdot k$ and $(p-k)^2=(P-q)^2=4(1-h-q^0)$ give $m_{-1}$, and the next order gives $m_0$, as exact rational functions. (ii) The Pauli components $c_0,c_n,c_r,c_k$ of $m_0$, defined from the two-spin amplitude $\psi$ by $X^T=\psi\,(i\sigma_y)$ (see the paragraph on the two-spin state below) in the basis $\{I,\sigma_n,\sigma_r,\sigma_k\}$, are contracted with the color Gram matrix and the polarization sum $\delta_{ij}-\hat\ell_i\hat\ell_j$, and projected onto the two parity sectors. Explicitly, for a color ordering $\sigma\in\{T^aT^b,T^bT^a\}$, outgoing-gluon polarization $\lambda$ and incoming-gluon polarization $\lambda_p\in\{\hat x,\hat y\}$, and with $\langle X|Y\rangle=\frac12{\rm Tr}(X^\dagger Y)$, Eqs.~\eqref{eq:Sigma} and \eqref{eq:Kdef} read
\begin{equation}
K_{AB}=E^2\int_\Omega\frac{d\Omega}{4\pi}\,
\frac{\sum_{\lambda,\lambda_p}\sum_{\sigma\sigma'}G_{\sigma\sigma'}\,\langle\widehat T_A|m_0^{\sigma}\rangle\,\langle m_0^{\sigma'}|\widehat T_B\rangle}{N_cC_F\,\langle X_0|X_0\rangle},
\qquad
\langle I|m_0\rangle=c_0,\quad\langle\sigma_n|m_0\rangle=c_n,\quad
\langle\widehat T_n|m_0\rangle=\frac{Ac_k-Bc_r}{\sqrt{A^2+B^2}},
\label{eq:chain}
\end{equation}
with $G$ the color Gram matrix of Eq.~\eqref{eq:Gram} and $N_cC_F={\rm Tr}(T^aT^a)$ the color factor of the Born amplitude, which is nonzero only for $\lambda_p=\hat x$. The initial-state averages are common to numerator and denominator and cancel. Then $K_n=K_{TT}$, $K_\perp=K_{II}+K_{nn}$, and $\Gamma_{n,\perp}=(\alpha_s/\pi)\tau_c^2K_{n,\perp}$. In the mirror-odd sector the densities of $K_\perp$, $K_t$ and $K_a$ are in the ratio $(|c_0|^2+|c_n|^2):2(|c_0|^2-|c_n|^2):-2\,{\rm Im}(c_0c_n^*)$, with the color and polarization sums of Eq.~\eqref{eq:chain} understood. In the resulting densities the factor $p\cdot\hat\ell\propto1+\hat\ell_z$ cancels identically, and the denominators reduce to $(1-\beta x)^2(1+\beta x)^2$ with $x=\hat\ell\cdot\hat k$. (iii) Parametrizing $\hat\ell$ as in Sec.~II, with $x=\cos\chi$, and averaging over the azimuth of $\hat\ell$ about $\hat k$ leaves the densities of Eq.~\eqref{eq:Kmaster}. With $c=\cos\theta$, $s=\sin\theta$, $D=1-\beta^2c^2$ and $\bar m\equiv\sqrt{1-\beta^2}$, their coefficients are
\begin{align}
n^\perp_0&=\frac{27-9c^2+\beta^2(9-24c^2+11c^4)-7\beta^4c^2(1+c^2)}{24\beta^2s^2D},\qquad
n^\perp_4=-\frac{7-21c^2+\beta^2(21-40c^2+15c^4)-9\beta^4c^2(3-5c^2)}{24s^2D},\nonumber\\
n^\perp_2&=\frac{-9+27c^2-\beta^2(6+79c^2+33c^4)+\beta^4(7+13c^2+98c^4)-9\beta^6c^2(1+c^2)}{24\beta^2s^2D},\qquad n^\perp_1=n^\perp_3=0,\nonumber\\
n^n_0&=\frac{\bar m^2(3-c^2)(9+7\beta^2c^2)}{24D^2},\qquad
n^n_4=\frac{\beta^2\bar m^2(3c^2-1)(7+9\beta^2c^2)}{24D^2},\nonumber\\
n^n_2&=-\frac{\bar m^2\big[9-27c^2-\beta^2(21-142c^2+21c^4)-9\beta^4c^2(3-c^2)\big]}{24D^2},\nonumber\\
n^n_1&=\frac{c\,\bar m^2(3-8\beta^2+5\beta^2c^2)}{2D^2},\qquad
n^n_3=\frac{\beta^2c\,\bar m^2(5-8c^2+3\beta^2c^2)}{2D^2},\nonumber\\
n^t_0&=\frac{27-9c^2-\beta^2(63-84c^2+25c^4)-7\beta^4c^2(5-3c^2)}{12\beta^2s^2D},\qquad
n^t_4=-\frac{7-21c^2-\beta^2(35-100c^2+69c^4)+9\beta^4c^2(1+c^2)}{12s^2D},\nonumber\\
n^t_2&=\frac{-9+27c^2+\beta^2(66-259c^2+75c^4)-\beta^4(49-325c^2+158c^4)-9\beta^6c^2(5-3c^2)}{12\beta^2s^2D},\qquad n^t_1=n^t_3=0,\nonumber\\
n^a_0&=-\frac{\bar m(18-9c^2+7\beta^2c^2)}{12\beta sD},\qquad
n^a_4=\frac{\beta\bar m(14-21c^2-9\beta^2c^2)}{12sD},\nonumber\\
n^a_2&=\frac{\bar m(18-27c^2-14\beta^2+64\beta^2c^2-9\beta^4c^2)}{12\beta sD},\nonumber\\
n^a_1&=\frac{c\,\bar m(3-8\beta^2+5\beta^2c^2)}{2\beta sD},\qquad
n^a_3=\frac{\beta c\,\bar m(5-8c^2+3\beta^2c^2)}{2sD}.
\label{eq:ncoef}
\end{align}
The odd coefficients do not contribute over the full solid angle but enter the cone corrections. (iv) Inserting Eq.~\eqref{eq:ncoef} into Eq.~\eqref{eq:Kmaster} with the integrals \eqref{eq:Lbeta} gives Eqs.~\eqref{eq:Kclosed}--\eqref{eq:Kpolys}. All steps were carried out in exact rational arithmetic, and $h$ cancels identically from the projected densities. As independent checks, the densities agree with the floating-point evaluation of the amplitude described in Appendix~B to ten digits, and the closed forms agree with a direct numerical integration of the exact densities to $10^{-14}$ at the points $(\beta,\theta)=(0.3,0.4)$, $(0.6,\pi/2)$, $(0.85,1.0)$, $(0.5,2.3)$ and $(0.95,0.2)$.

\paragraph*{Two-spin state and rotations.} The antiquark Pauli spinor of Eq.~(6) is related to the physical antiquark spin state $\chi$ by $\eta=-i\sigma_y\chi^*$. The two-spin wave function in the physical spin basis, $\psi_{\lambda_q\lambda_{\bar q}}$, is therefore related to the operator of Eq.~(6) by $\psi\,(i\sigma_y)=X^T$, and the Hilbert-Schmidt product ${\rm Tr}(X^\dagger Y)$ equals the overlap of the corresponding two-spin states. This is the correspondence $|\widehat T_A\rangle\leftrightarrow\widehat T_A$ used in Eq.~\eqref{eq:drho}. A common rotation of both spins, $\psi\to U\psi U^T$ with $U\in SU(2)$, acts as $X^T\to UX^TU^\dagger$, since $\sigma_yU^T\sigma_y=U^\dagger$. For a rotation about $\hat n=\hat y$, $U=e^{-i\delta\phi_n\sigma_n/2}$ is real, so $X\to UXU^\dagger$. This is the active rotation of Eq.~\eqref{eq:active}, and the displacement $\delta\phi_n T_n$ is the tangent of the physical rotation of the two-spin state by $\delta\phi_n$ about $\hat n$. In the same way, $\Pi_n$ acts as $X\to-\sigma_nX\sigma_n$, which leaves $\sigma_r,\sigma_k$ invariant and reverses $I$ and $\sigma_n$.

\paragraph*{Variation of the spin basis.} For a general azimuth $\varphi$ of the quark direction about the photon axis,
$\hat k=(\sin\theta\cos\varphi,\sin\theta\sin\varphi,\cos\theta)$ and $\hat r=(\cos\theta\cos\varphi,\cos\theta\sin\varphi,-\sin\theta)$. At $\varphi=0$,
\begin{equation}
\delta\hat k=\delta\theta\,\hat r+\sin\theta\,\delta\varphi\,\hat n,
\qquad
\delta\hat r=-\delta\theta\,\hat k+\cos\theta\,\delta\varphi\,\hat n ,
\end{equation}
and therefore $\delta\sigma_k=\delta\theta\,\sigma_r+\sin\theta\,\delta\varphi\,\sigma_n$ and $\delta\sigma_r=-\delta\theta\,\sigma_k+\cos\theta\,\delta\varphi\,\sigma_n$. Inserting these into $\delta X_0=\delta A\,\sigma_r+\delta B\,\sigma_k+A\,\delta\sigma_r+B\,\delta\sigma_k$ gives Eq.~\eqref{eq:delta_X0_theta_collect}. Projecting onto $T_n$ with ${\rm Tr}(\sigma_i\sigma_j)=2\delta_{ij}$,
\begin{equation}
\frac{{\rm Tr}\big[T_n^\dagger\delta X_0^{(\theta)}\big]}{{\rm Tr}\big[T_n^\dagger T_n\big]}
=\left[\frac{AB_\theta-BA_\theta}{A^2+B^2}-1\right]\delta\theta .
\end{equation}
Here $-\delta\theta$ comes from the transport of the axes, whose assignment depends on the spinor basis (Sec.~\ref{sec:intrinsic}), and the first term is the connection \eqref{eq:dphitheta_result}. The $\sigma_n$ term, proportional to $\delta\varphi$, lies in the $\Pi_n=-1$ sector.

\paragraph*{Intrinsic-spin projection.} With $X_0=\boldsymbol\sigma\cdot\mathbf x$, $T_n=\boldsymbol\sigma\cdot\mathbf t$, $\mathbf x=A\hat r+B\hat k$ and $\mathbf t=A\hat k-B\hat r=\mathbf x\times\hat n$ (the triad of Eq.~(2) gives $\hat n\times\hat r=-\hat k$ and $\hat n\times\hat k=\hat r$), the identity $(\mathbf a\cdot\boldsymbol\sigma)(\mathbf x\cdot\boldsymbol\sigma)=\mathbf a\cdot\mathbf x\,I+i\boldsymbol\sigma\cdot(\mathbf a\times\mathbf x)$ and the decomposition $\mathbf a\times\mathbf x=[(\mathbf a\times\mathbf x)\cdot\hat n]\,\hat n-a_n\,\mathbf t$ give Eq.~\eqref{eq:spin-insertion}. The tangent projection is $-i\,a_n$ for a quark-side insertion and $+i\,b_n$ for an antiquark-side insertion. The opposite signs reflect the left and right action of the two insertions on the Pauli operator.

\paragraph*{Charge-two transformation.} Under independent rotations of the quark and antiquark spins by $\phi_q,\phi_{\bar q}$ about $\hat n$, $C_\parallel\to R(\phi_q)C_\parallel R^T(\phi_{\bar q})$. Writing $C_\parallel=\frac{t}{2}I+aJ+\begin{pmatrix}q_1&q_2\\q_2&-q_1\end{pmatrix}$ with $J=\begin{pmatrix}0&-1\\1&0\end{pmatrix}$, the combination $t+2ia$ acquires the phase of $\phi_q-\phi_{\bar q}$ and ${\cal Q}=q_1+iq_2$ that of $\phi_q+\phi_{\bar q}$. For a common rotation ${\cal Q}\to e^{2i\phi}{\cal Q}$, and for a Gaussian, zero-mean $\Delta\phi$ one has $\langle e^{2i\Delta\phi}\rangle=e^{-2\langle\Delta\phi^2\rangle}$, which is the factor used in Eq.~\eqref{eq:exp-n}.

\section{Appendix B: Amplitudes leading to subleading Soft theorem}
\label{app:amplitudes}

We write the fermion propagator as $S(\ell)=i(\slashed\ell+m)/(\ell^2-m^2)$, with $\ell$ the momentum flowing along the fermion line from $v(p_2)$ to $\bar u(p_1)$, and use $q+p=p_1+p_2+k$. The eight tree-level diagrams for $\gamma^*(q)+g(p,a)\to q(p_1)+\bar q(p_2)+g(k,b)$ are
\begin{eqnarray}\nonumber
i{\mathcal M}_1&=&\bar u(p_1)\left(i g_s \gamma^\rho T^b\right)S(p_1+k)\left(i e e_q \gamma^\mu\right)S(p_1+k-q)\left(i g_s \gamma^\nu T^a\right)v(p_2)\,\varepsilon_\mu(q)\,\varepsilon_\nu(p)\,\varepsilon_\rho^{*}(k),\\ \nonumber
i{\mathcal M}_2&=&\bar u(p_1)\left(i e e_q\gamma^\mu\right)S(p_1-q)\left(i g_s \gamma^\rho T^b\right)S(p_1+k-q)\left(i g_s \gamma^\nu T^a\right)v(p_2)\,\varepsilon_\mu(q)\,\varepsilon_\nu(p)\,\varepsilon_\rho^{*}(k),\\ \nonumber
i{\mathcal M}_3&=&\bar u(p_1)\left(i e e_q \gamma^\mu\right)S(p_1-q)\left(i g_s \gamma^\nu T^a\right)S(-p_2-k)\left(i g_s \gamma^\rho T^b\right)v(p_2)\,\varepsilon_\mu(q)\,\varepsilon_\nu(p)\,\varepsilon_\rho^{*}(k),\\ \nonumber
i{\mathcal M}_4&=&\bar u(p_1)\left(i g_s \gamma^\rho T^b\right)S(p_1+k)\left(i g_s \gamma^\nu T^a\right)S(p_1+k-p)\left(i e e_q \gamma^\mu\right)v(p_2)\,\varepsilon_\mu(q)\,\varepsilon_\nu(p)\,\varepsilon_\rho^{*}(k),\\ \nonumber
i{\mathcal M}_5&=&\bar u(p_1)\left(i g_s \gamma^\nu T^a\right)S(p_1-p)\left(i g_s \gamma^\rho T^b\right)S(p_1+k-p)\left(i e e_q \gamma^\mu\right)v(p_2)\,\varepsilon_\mu(q)\,\varepsilon_\nu(p)\,\varepsilon_\rho^{*}(k),\\ \nonumber
i{\mathcal M}_6&=&\bar u(p_1)\left(i g_s \gamma^\nu T^a\right)S(p_1-p)\left(i e e_q \gamma^\mu\right)S(-p_2-k)\left(i g_s \gamma^\rho T^b\right)v(p_2)\,\varepsilon_\mu(q)\,\varepsilon_\nu(p)\,\varepsilon_\rho^{*}(k),\\ \nonumber
i\mathcal{M}_7 &=&\bar{u}(p_1)\left(i e e_q \gamma^\mu\right)S(p_1-q)\left(i g_s T^c\gamma^\sigma\right)v(p_2)\,
\frac{-i}{(p-k)^2}\left(i g_s f^{abc}\right)V_{\nu\rho\sigma}\big(p,-k,-(p-k)\big)\,\varepsilon_\mu(q)\,\varepsilon_\nu(p)\,\varepsilon^{*}_{\rho}(k),\\ \nonumber
i\mathcal{M}_8 &=&\bar{u}(p_1)\left(ig_s T^c\gamma^\sigma \right)S(p_1+k-p)\left(iee_q \gamma^\mu \right)v(p_2)\,
\frac{-i}{(p-k)^2}\left(i g_s f^{abc}\right)V_{\nu\rho\sigma}\big(p,-k,-(p-k)\big)\,\varepsilon_\mu(q)\,\varepsilon_\nu(p)\,\varepsilon^{*}_{\rho}(k),
\end{eqnarray}
with the three-gluon vertex for incoming momenta $(p,-k,-(p-k))$,
\begin{align}
V_{\nu\rho\sigma}\big(p,-k,-(p-k)\big)
&=
-\Big[(p+k)_{\sigma} g_{\nu\rho}+(p-2k)_{\nu} g_{\rho\sigma}+(k-2p)_{\rho} g_{\sigma\nu}\Big] .
\label{eq:V3}
\end{align}
The Born amplitude is ${\cal M}_{\rm Born}={\cal M}_A+{\cal M}_B$, with
\begin{eqnarray}
i{\cal M}_A&=&\bar u(p_1)\left(iee_q\gamma^\mu\right)S(p_1-q)\left(ig_sT^a\gamma^\nu\right)v(p_2)\,\varepsilon_\mu(q)\varepsilon_\nu(p),\nonumber\\
i{\cal M}_B&=&\bar u(p_1)\left(ig_sT^a\gamma^\nu\right)S(p_1-p)\left(iee_q\gamma^\mu\right)v(p_2)\,\varepsilon_\mu(q)\varepsilon_\nu(p).
\label{eq:Born-AB}
\end{eqnarray}
Each diagram has the soft expansion $i{\cal M}_j=\omega^{-1}i{\cal M}^{(-1)}_j+i{\cal M}^{(0)}_j+{\cal O}(\omega)$.

\subsection{External legs: eikonal and intrinsic-spin terms}

For emission from an external heavy-quark line the Dirac equation gives the exact identities
\begin{equation}
\bar u(p_1)\gamma^\rho(\slashed p_1+\slashed k+m)=\bar u(p_1)\left(2p_1^\rho+\gamma^\rho\slashed k\right),
\qquad
(-\slashed p_2-\slashed k+m)\gamma^\rho v(p_2)=-\left(2p_2^\rho+\slashed k\gamma^\rho\right)v(p_2),
\label{eq:dirac-ext}
\end{equation}
with $(p_{1,2}+k)^2-m^2=2p_{1,2}\cdot k$. With $S^{\mu\nu}=\frac14[\gamma^\mu,\gamma^\nu]$ and $\varepsilon^*\cdot k=0$, one has $\slashed\varepsilon^*\slashed k=2\varepsilon^*_\mu k_\nu S^{\mu\nu}$ and $\slashed k\slashed\varepsilon^*=-2\varepsilon^*_\mu k_\nu S^{\mu\nu}$, so that, without any expansion,
\begin{align}
\bar u(p_1)\left(ig_s\gamma^\rho T^b\right)S(p_1+k)\,\varepsilon^*_\rho
&=-g_s\,T^b\,\bar u(p_1)\left[\frac{p_1\cdot\varepsilon^*}{p_1\cdot k}+\frac{\varepsilon^*_\mu k_\nu S^{\mu\nu}}{p_1\cdot k}\right],
\nonumber\\
S(-p_2-k)\left(ig_s\gamma^\rho T^b\right)v(p_2)\,\varepsilon^*_\rho
&=+g_s\left[\frac{p_2\cdot\varepsilon^*}{p_2\cdot k}-\frac{\varepsilon^*_\mu k_\nu S^{\mu\nu}}{p_2\cdot k}\right]T^b\,v(p_2).
\label{eq:ext-exact}
\end{align}
The first term in each bracket is the eikonal factor and the second the intrinsic-spin insertion of Eq.~\eqref{eq:soft-theorem}; the opposite signs of the spin terms reflect the action of the generator on $\bar u$ from the right and on $v$ from the left. The remainder of diagrams 1 and 4 is the Born amplitude ${\cal M}_A$, respectively ${\cal M}_B$, with $p_1\to p_1+k$, and the remainder of diagrams 3 and 6 is the Born amplitude with $p_2\to p_2+k$. Hence, in the limit $\omega\to0$,
\begin{equation}
{\cal M}^{(-1)}_{1+4}=-g_s\,T^b\,\frac{p_1\cdot\varepsilon^*}{p_1\cdot k}\,{\cal M}_{\rm Born},
\qquad
{\cal M}^{(-1)}_{3+6}=+g_s\,\frac{p_2\cdot\varepsilon^*}{p_2\cdot k}\,{\cal M}_{\rm Born}\,T^b ,
\label{eq:eik-quarks}
\end{equation}
where $T^b$ multiplies the color matrix of the Born amplitude from the left or from the right, and at the next order the external diagrams give
\begin{equation}
{\cal M}^{(0)}_{1+3+4+6}=g_s\left[-T^b\frac{p_1\cdot\varepsilon^*}{p_1\cdot k}\,k\cdot\partial_{p_1}{\cal M}_{\rm Born}+\frac{p_2\cdot\varepsilon^*}{p_2\cdot k}\,k\cdot\partial_{p_2}{\cal M}_{\rm Born}T^b
-T^b\frac{\varepsilon^*_\mu k_\nu}{p_1\cdot k}S^{\mu\nu}_1{\cal M}_{\rm Born}-\frac{\varepsilon^*_\mu k_\nu}{p_2\cdot k}S^{\mu\nu}_2{\cal M}_{\rm Born}T^b\right],
\label{eq:ext-nts}
\end{equation}
with $S_{1}$ ($S_2$) the generator $S^{\mu\nu}$ acting on the spinor index of $\bar u(p_1)$ ($v(p_2)$) as in Eq.~\eqref{eq:ext-exact}. Diagrams 2 and 5, in which the gluon is emitted from the internal line, are finite as $\omega\to0$ and do not contribute to the eikonal term.

\subsection{Internal emission and the Ward identity}

For the internal line, $S(\ell)\left(ig_s\gamma^\rho T^b\right)S(\ell+k)=g_s\,T^b\,\partial S(\ell)/\partial\ell_\rho+{\cal O}(k)$, since $\partial(\slashed\ell-m)^{-1}/\partial\ell_\rho=-(\slashed\ell-m)^{-1}\gamma^\rho(\slashed\ell-m)^{-1}$. Diagrams 2 and 5 therefore give
\begin{equation}
{\cal M}^{(0)}_{2+5}=g_s\left[T^b\,\varepsilon^*_\alpha\,\frac{\partial{\cal M}_A}{\partial p_{1\alpha}}\bigg|_{\rm int}-\varepsilon^*_\alpha\,\frac{\partial{\cal M}_B}{\partial p_{2\alpha}}\bigg|_{\rm int}T^b\right],
\label{eq:int-nts}
\end{equation}
where the derivatives act on the internal propagators $S(p_1-q)$ of ${\cal M}_A$ and $S(p_1-p)=S(q-p_2)$ of ${\cal M}_B$ at fixed $q$ and $p$. Replacing $\varepsilon^*\to k$ in the sum of all diagrams must give zero. The eikonal terms then give $\mp g_sT^b{\cal M}_{\rm Born}$ terms that cancel against the incoming-gluon term by color conservation, and at the next order the derivative terms of Eqs.~\eqref{eq:ext-nts} and \eqref{eq:int-nts} must combine into the gauge-invariant structure
\begin{equation}
\frac{(p_i\cdot\varepsilon^*)\,k^\alpha-(p_i\cdot k)\,\varepsilon^{*\alpha}}{p_i\cdot k}\,\frac{\partial}{\partial p_i^\alpha}
=\frac{\varepsilon^*_\mu k_\nu L_i^{\mu\nu}}{p_i\cdot k},
\label{eq:L-combine}
\end{equation}
which vanishes for $\varepsilon^*\to k$. Together with the spin terms of Eq.~\eqref{eq:ext-nts}, the heavy-quark contributions are
\begin{equation}
{\cal M}^{(0)}=-g_s\sum_{i=1,2}{\bf T}_i^b\,\frac{\varepsilon^*_{\mu} k_\nu J_i^{\mu\nu}}{p_i\cdot k}\,{\cal M}_{\rm Born},
\qquad J_1^{\mu\nu}=L_1^{\mu\nu}+S_1^{\mu\nu},\quad J_2^{\mu\nu}=L_2^{\mu\nu}-S_2^{\mu\nu},
\label{eq:LBK-final}
\end{equation}
in the all-outgoing color notation of Eq.~\eqref{eq:soft-theorem}, where ${\bf T}_1^b$ acts as $T^b$ from the left and ${\bf T}_2^b$ as $-T^b$ from the right, and where the spin part of $J_2$ is $-S_2$, as follows from Eq.~\eqref{eq:ext-exact}: the generator acting on $v(p_2)$ enters with the opposite sign. The overall sign relative to Eq.~\eqref{eq:soft-theorem} is a common phase convention for the emission vertex and drops out of the density matrix. The emissions from the incoming gluon, diagrams 7 and 8, give the eikonal term ${\cal M}^{(-1)}_{7+8}\propto g_sf^{abc}\,(p\cdot\varepsilon^*/p\cdot k)\,{\cal M}^c_{\rm Born}$ and the corresponding next-to-soft term with $p\cdot k$ in the denominator, which completes the sum over all colored legs. In the complete evaluation of Sec.~\ref{sec:kernels} and of the following subsection, the next-to-soft amplitude is obtained by expanding all eight diagrams, not from Eq.~\eqref{eq:LBK-final}, and the denominator $p\cdot k$ cancels in the projected densities.

\subsection{Evaluation of the complete next-to-soft amplitude}
\label{app:numerics}

The closed forms of Sec.~\ref{sec:kernels} follow from the exact expansion of the diagrams above, as described in Appendix~A. The cone corrections $\Delta K$ and the benchmark values of Sec.~\ref{sec:eic_observable} require an integration over a region without azimuthal symmetry about $\hat k$. For these we evaluate the same eight diagrams (the six orderings of the photon and the two gluons along the heavy-quark line and the two three-gluon diagrams) directly with Dirac matrices, with canonical heavy-quark spinors defined by boosts from the pair rest frame and the antiquark spin axes $\{\hat n,-\hat r,-\hat k\}$. The kinematics is fixed by the observed $(M,\beta,\theta)$, by $Q^2$, and by $q+p=p_1+p_2+k$, $q^2=-Q^2$, $p^2=0$, with the photon along $+\hat z$ in the pair rest frame. The incoming-gluon polarization is averaged over $\hat x$, in the production plane, and $\hat y$, normal to it, and its color over the adjoint index; the same averages appear in the Born norm and cancel in Eq.~\eqref{eq:chain}. With $f^{abc}T^c=-i[T^a,T^b]$ the amplitude has two color-ordered components, and the color Gram matrix of the two orderings is
\begin{equation}
G=\begin{pmatrix}N_cC_F^2&-C_F/2\\-C_F/2&N_cC_F^2\end{pmatrix}.
\label{eq:Gram}
\end{equation}
This floating-point evaluation is independent of the exact expansion of Appendix~A, and the following checks were performed on it for sampled gluon directions and polarizations at several values of $(\beta,\theta)$; checks (i)--(iv) hold to $10^{-11}$ or better. (i) At $\omega=0$ the amplitude reproduces Eqs.~\eqref{eq:crr}--\eqref{eq:crk} and $C_{nn}=1$. (ii) Both color-ordered components vanish under $\epsilon(k)\to k$ and under $\epsilon(p)\to p$, which fixes the relative normalization of the three-gluon diagrams. (iii) $m_{-1}$ is proportional to $X_0$ for every direction, color ordering and polarization, and its normalized square with $G$ reproduces the dipole kernel \eqref{eq:W0}. (iv) The off-diagonal elements $C_{nr},C_{nk}$ and the single-spin polarizations vanish, as required by Eq.~\eqref{eq:blockdiag} and by the absence of absorptive phases. (v) The kernels are independent of $Q^2$ at fixed $(\beta,\theta)$ for $0.01\le Q^2/M^2\le4$, in agreement with the exact result of Appendix~A. Here $m_{-1}$ and $m_0$ are extracted from symmetric combinations of the amplitude at $\pm\omega$ and $\pm2\omega$ with $\omega=10^{-3}E$; the resulting densities agree with the exact rational densities of Appendix~A to ten digits. The integration over $\Omega_{\rm wa}$ uses a midpoint grid aligned with the cones; its discretization error is about $1\%$ of $K$.

\bibliography{ref.bib}

\end{document}